\documentclass{article}

\usepackage{arxiv}

\usepackage[utf8]{inputenc} 
\usepackage[T1]{fontenc}    
\usepackage{hyperref}       
\usepackage{url}            
\usepackage{booktabs}       
\usepackage{amsfonts}       
\usepackage{nicefrac}       
\usepackage{microtype}      
\usepackage{lipsum}		
\usepackage{graphicx}
\usepackage{doi}

\usepackage{subfiles}
\usepackage[acronym, toc]{glossaries}
\usepackage[ruled,vlined]{algorithm2e}
\usepackage{tabularx}
\usepackage{caption}
\usepackage{graphicx}
\usepackage{ragged2e}
\usepackage{multirow}
\usepackage{amsmath}
\usepackage{geometry}
\usepackage{float}
\usepackage[noend]{algpseudocode}
\usepackage{appendix}
\usepackage[figuresright]{rotating}

\graphicspath{ {./Figures/} }

\newcommand{\var}{\texttt}
\newcommand{\proc}[2]{\textsl{#1}(#2)}

\title{Gotta catch 'em all: a Multistage Framework for honeypot fingerprinting}

\author{{\hspace{1mm}Shreyas Srinivasa} \\
	Department of Electronic Systems\\
	Aalborg University\\
	Copenhagen, 2450 \\
	\texttt{shsr@es.aau.dk} \\
	\And
	Jens Myrup Pedersen\\
	Department of Electronic Systems\\
	Aalborg University\\
	Copenhagen, 2450 \\
	\texttt{jens@es.aau.dk} \\
	\And
    Emmanouil Vasilomanolakis \\
	Department of Electronic Systems\\
	Aalborg University\\
	Copenhagen, 2450 \\
	\texttt{emv@es.aau.dk} \\
}

\date{}

\renewcommand{\shorttitle}{a Multistage Framework for honeypot fingerprinting}

\newacronym{isp}{ISP}{Internet Service Provider}
\newacronym{ids}{IDS}{Intrusion Detection Systems}
\newacronym{ics}{ICS}{Industrial Control Systems}
\newacronym{as}{AS}{Autonomous Systems}
\newacronym{tcp}{TCP}{Transmission Control Protocol}
\newacronym{icw}{ICW}{Initial Congestion Window}
\newacronym{rto}{RTO}{Retransmission Time Out}
\newacronym{dns}{DNS}{Domain Name Service}
\newacronym{fqdn}{FQDN}{Fully Qualified Domain Name}
\newacronym{tls}{TLS}{Transport Layer Security}
\newacronym{ip}{IP}{Internet Protocol}
\newacronym{api}{API}{Application Programming Interface}
\newacronym{it}{IT}{Information Technology}
\newacronym{vpn}{VPN}{Virtual Private Network}
\newacronym{cdn}{CDN}{Content Delivery Network}
\newacronym{ddos}{DDoS}{Distributed Denial of Service}
\newacronym{os}{OS}{Operating System}
\newacronym{mtd}{MTD}{Moving Target Defense}
\newacronym{nat}{NAT}{Network Address Translation}

\begin{document}
\maketitle

\begin{abstract}
Honeypots are decoy systems that lure attackers by presenting them with a seemingly vulnerable system. They provide an early detection mechanism as well as a method for learning how adversaries work and think.
However, over the last years a number of researchers have shown methods for fingerprinting honeypots. This significantly decreases the value of a honeypot; if an attacker is able to recognize the existence of such a system, they can evade it. 
In this article, we revisit the honeypot identification field, by providing a holistic framework that includes state of the art and novel fingerprinting components. 
We decrease the probability of false positives by proposing a rigid multi-step approach for labeling a system as a honeypot.
We perform extensive scans covering $2.9$ billion addresses of the IPv4 space and identify a total of $21,855$ honeypot instances.
Moreover, we present a number of interesting side-findings such as the identification of more than $354,431$ non-honeypot systems that represent potentially vulnerable servers (e.g. SSH servers with default password configurations and vulnerable versions).
Lastly, we discuss countermeasures against honeypot fingerprinting techniques.
\end{abstract}

\keywords{Honeypots \and Honeypot Fingerprinting}

 \section{Introduction}

Honeypots are decoy systems whose only value lies in being probed, attacked, and compromised. They attempt to lure attackers in, to provide an early warning system, and act as a method for understanding the adversaries' mindset and determining new attack trends \cite{Spitzner03}. 
Honeypots are not a stand alone security mechanism but rather important supplements to existing infrastructure (e.g. firewalls and \acrfull{ids}). Nevertheless, they offer a unique attack understanding and perspective, while exhibiting a very low number of false positives.
The latter is due to the fact that any communication towards a honeypot is considered hostile; i.e. benign users have no reason to contact a honeypot. 
%

Honeypots are commonly classified based on the interaction level they offer to the adversary. This results in low-, medium- and high- interaction honeypots \cite{nawrocki2016survey}. While the first two categories offer different levels of \textit{emulation} of protocols, the latter (i.e. high-interaction) describes real world systems. High-interaction honeypots are too expensive to maintain and significantly less used than low/medium interaction; hence, we consider them out of the scope of this article. 
Over the years, low and medium interaction honeypots have been designed and developed to emulate the majority of commonly used protocols. These include SSH (e.g. Kippo \cite{KIPPO} and Cowrie \cite{oosterhof2016cowrie}), Telnet (e.g. Cowrie \cite{oosterhof2016cowrie}), HTTP (e.g. Glastopf \cite{rist2009glastopf}), FTP, SMB (e.g. Dionaea \cite{Dionaea} and HosTaGe \cite{hostage_generic}) and also \acrfull{ics} protocols like Modbus and S7 (e.g. Conpot \cite{rist2013conpot} and HosTaGe \cite{hostage}). 

One of the key success criteria for a honeypot is that it is indistinguishable from a real system. This can be translated to the following axiom: \textit{if a honeypot can be easily identified as  such, then its value is significantly decreased}. The reason for this is that an adversary can potentially either evade honeypots (e.g. perform reconnaissance and add a blocklist of IP addresses into their malware, to avoid honeypots and reduce the risk of detection \cite{haware}) or attempt to take them down (e.g. via a \acrfull{ddos} attack). Note that modern malware (e.g. Hide’n Seek \cite{bitdefender}) already include hard-coded IP addresses (e.g. belonging to known security agencies) that are blocklisted from all communications.  
Honeypot fingerprinting is the process of revealing that a seemingly vulnerable system is, in fact, a honeypot.

In this article, we perform a comprehensive analysis of honeypot fingerprinting techniques. For this, we present a holistic framework that includes a number of novel fingerprinting methods along with all major state of the art techniques. 
Among others, we propose a new protocol handshake fingerprinting component, a static \acrfull{tls} certificate method and a \acrfull{fqdn} check. Furthermore, we present the results of extensive honeypot identification scans over the Internet for $9$ prominent honeypot implementations. 
Our results come as an independent confirmation of previous studies (\cite{Vetterl2018,morishita}) but also as a step forward to a more holistic study of honeypots. In particular, due to the multistage checks that our framework performs, we argue that the presented results have a very low probability for false positives. Moreover, we present several insights for IP addresses that are not marked as honeypots, but are likely to be real vulnerable systems. Lastly, we discuss ethical considerations and possible countermeasures against fingerprinting. 
The core contributions of this article can be summarized as follows:
\begin{itemize}
    \item We present novel methods for active honeypot fingerprinting (so-called probe-based). These are combined with a number of SotA and third-party (so-called meta-scan) fingerprinting techniques in the form of a multistage fingerprinting framework. We scan $2.9$ billion IP addresses of the IPv4 space, discover  $187$ million IP addresses with relevant open ports and identify \textbf{a total of 21,855 honeypots}.
   
    \item We showcase that out of the 21,855 identified honeypots, third-party techniques can only reveal $33.9\%$ of the total honeypot population. On the contrary, we show that $100\%$ of the honeypots can be detected via our probe-based methodology.  
   
    \item As a side finding, we identify more than $260,000$ non-honeypot entities (i.e. SSH and FTP servers) that appear to use trivial passwords and/or are susceptible to high-severity vulnerabilities. 
    
\end{itemize}

The rest of the article is structured as follows. We propose our framework for honeypot fingerprinting in Section \ref{sec:multistage}. Section \ref{sec:eval} presents our evaluation. Section \ref{sec:discussion} discusses ethical considerations, fingerprinting countermeasures and the limitations. Section \ref{sec:rw} presents the related work on honeypot fingerprinting research. We conclude the article in Section \ref{sec:conc}.

\section{Multistage Honeypot Fingerprinting Framework}
\label{sec:multistage}

    Researchers classify fingerprinting techniques as active and passive, based on attacker-honeypot interaction \cite{spitzner}. Active-fingerprinting involves creating specific probes and using them to querying the target system to collect as much data possible. On the contrary, passive-fingerprinting makes use of available data about the target system for further analysis to determine information about the target. 
    
    In the following, we attempt to examine methods in both the active and passive spectrum in Section \ref{subsec:overview} . On the one hand, we assume that attackers prefer passive methods since they come with multiple benefits. Mainly, they are stealthier (i.e. no direct communication to the honeypot is needed) and easier to use (e.g. systems like Shodan \cite{SHODAN} already exist and offer an \acrfull{api} for such purposes). On the other hand, our hypothesis is that active approaches can identify a much broader set of honeypots. We propose the novel framework (see Section \ref{subsec:framework}) that utilizes both active and passive fingerprinting techniques to fingerprint honeypots deployed on the Internet. The aim of the proposed framework is to systematically fingerprint honeypots with multiple sequential checks to reduce false positives. In comparison to state of the art (c.f. Section \ref{sec:rw}), we employ novel probing methods that include certificate checks, protocol handshake and metascan methods that check for  \acrfull{isp} and cloud hosting information. The framework is further automated for all the checks involved in each fingerprinting technique that helps in automated transition to stages during the scanning process.  

\subsection{Overview}
\label{subsec:overview}
This section provides an overview of the proposed multistage fingerprinting framework and the detection techniques.

\subsubsection{Probe-based fingerprinting}
Probe-based fingerprinting involves the creation of queries to derive fingerprinting information and involves direct interaction with a system. These methods focus on leveraging the data from responses and classifying the target machine based on fingerprinting identifiers. The information may include system-specific unique identifiers like the \acrfull{icw} or the \acrfull{rto}. Several fingerprinting tools like NMap \cite{NMap}, XProbe2 \cite{xprobe}, Metasploit \cite{metasploit}, and Hydra \cite{hydra} utilize probe-based methods to determine the \acrfull{os} and the protocol versions of the target systems. For example, these tools rely on banners advertised and the \textit{\acrfull{tcp}} header information to determine the underlying \acrshort{os}. The database of the scanning tool stores the identifiers that are specific to some \acrshort{os}.  The identifiers help compare parameter values obtained through probing for determining the \acrshort{os}. The fingerprinting probes derive multilevel system information at network-level, application-level, and the system-level. The integration of the information received from different levels improves detection accuracy.

\begin{figure}[t!]
    \centering
    \includegraphics[width=0.95\textwidth]{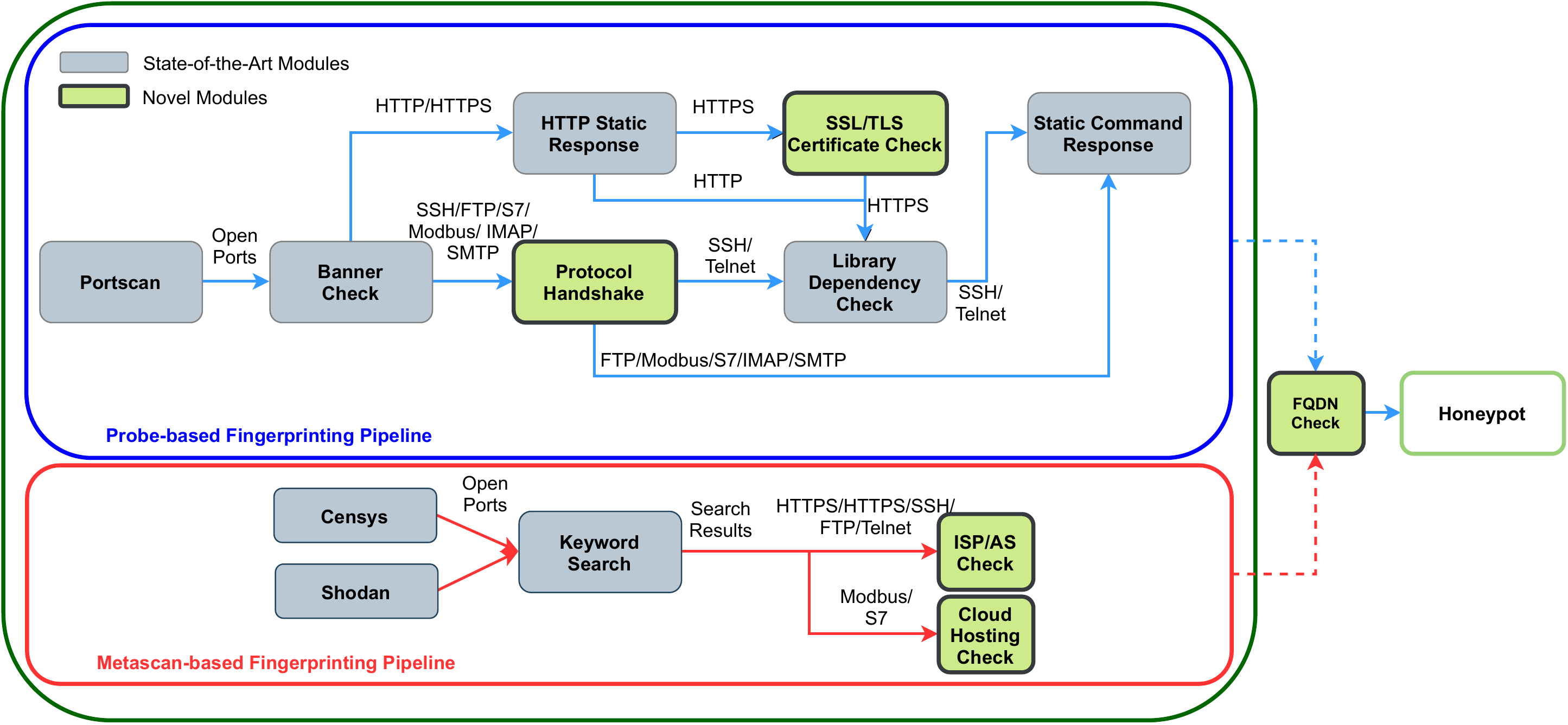}
    \caption{Multistage Framework for Honeypot Fingerprinting}
    \label{fig:fsm}
\end{figure}

\subsubsection{Metascan-based fingerprinting}
Metascan-based fingerprinting is a form of passive-fingerprinting that leverages the known information about the target system without direct interaction. The technique uses the \acrshort{ip} address and performs a search on Internet mass-scan engines (e.g. Shodan \cite{SHODAN} and Censys \cite{censys}) to obtain attributes like hosting provider and the \acrfull{isp}. The data obtained through metascan can be leveraged for fingerprinting purposes.
For example, if the target system has the \acrshort{tcp} port $502$ (i.e. the Modbus protocol default port) open  and the \acrshort{ip} address is attached with a network assigned to a university or research facility, this might act as an indication that the target system is a research honeypot. Similarly, if a cloud provider hosts the aforesaid system, it is likely a honeypot because \acrshort{ics} are physical devices that are deployed in an industrial network and are unlikely to be hosted by a cloud provider.

Mass-scan engines like Shodan and Censys crawl the Internet \acrshort{ip}-space daily to find vulnerable systems exposed to the Internet. They also store system- and network-specific information about the exposed systems like banners, \textit{HTTP} content, certificate, open ports, and services. Furthermore, they provide metadata like the \acrshort{isp}, \acrfull{as}, and geo-location of the systems.
Metascan fingerprinting techniques rely on essential information about target systems for the fingerprinting process. Such information can be obtained through the APIs offered by Shodan and Censys. Hence, the mass-scan engines can act as a substitute for the probe-based checks and provide the information required without interacting with the target systems.

\subsection{Framework}
\label{subsec:framework}
We construct a framework that combines both probe- and metascan-based methodologies. The framework is automated for the sequential checks for the probe-based and the metascan-based techniques. The probe-based technique uses methods that involve direct interaction with the target machine to fetch fingerprinting-information, while the metascan-based techniques use methods that involve no direct interaction with the target systems. In particular, the latter uses information derived from the Shodan and Censys mass-scan engines. Some mass-scan engines employ banner-based fingerprinting to fingerprint device types. For example, Censys \cite{censys} uses the Recog engine \cite{Recog} to detect device types using the information received by probing. The methods used in the proposed metascan pipeline are novel specifically towards honeypot fingerprinting. The novel methods of using the information about the ISP and checking if the instance is on a cloud environment assist us in gathering additional information that can be leveraged for the fingerprinting process. In addition, these methods help in reducing of false positives from the results. 
We term the target system considered for fingerprinting as an \textit{instance} for the rest of our article. 

Figure \ref{fig:fsm} shows the proposed multistage fingerprinting framework. The framework contains two independent main pipelines of \textit{Probe-based} and the \textit{Metascan-based} techniques. The \textit{Probe-based Pipeline} has multiple stages that are represented as boxes in the figure, with each stage aiming at fingerprinting the instance at various levels (i.e. network-level, application-level, system-level, protocol-level, implementation-level, and configuration-level). The boxes are color-coded to gray and green for further classification. The gray boxes denote that the stages refer to the state of the art, while the green boxes represent the novel methods. The novel methods are comprised of persistent checks that enrich the likelihood of the instance to be a honeypot. In the \textit{Metascan-based Pipeline}, the stages represent systematic checks referring to passive-fingerprinting techniques and data analysis. 
Overall, an \textit{instance} is only labeled as a \textit{honeypot} if all (relevant) components of the respective pipeline concur.    

\subsubsection{Probe-based Fingerprinting Pipeline}
\label{subsub: PBFP}
The Probe-based fingerprinting pipeline consists of \textit{seven} probing stages. The instances under evaluation transition into the next stage, based on the underlying application service protocol. The probes from each stage fetch information which is then analyzed to derive whether the instance is a honeypot. 

\paragraph{Portscan}
The pipeline begins by performing a scan on the Internet for open ports specific to the services emulated by the honeypots. Our framework utilizes ZMap \cite{zmap} for this process (alternatively one could use Masscan \cite{graham2014masscan}). The search results consist of a list of instances having these ports open to the Internet. 

\paragraph{Banner Check}
The results of the portscan are further processed in the banner check stage. In this stage, the probes check the banner advertised by the end system with static banners offered by honeypot implementations. Honeypot implementations offer a limited set of banners or even static banners that, in some cases, do not match the actual banners advertised by the services running on the underlying \acrshort{os}. As these banners are hard-coded, they can be matched against a list of known honeypot banners. We use the extended banner grab utility offered by ZMap to fetch banners from instances \cite{banner-grab}. SotA honeypot fingerprinting techniques by Vetterl et al. \cite{Vetterl2018} and Morishita et al. \cite{morishita} employ banner based fingerprinting to detect honeypots. We combine this knowledge (see Table \ref{tab:hp-banners} in the Appendix) to construct a holistic banner list for our framework. 
The results of this stage provide us with a list of instances and their banners. The instances that match the banners advertised by honeypots progress into the next stage based on the underlying protocol. For the instances that do not match the banners, we perform a vulnerability check that determines the number of vulnerable systems on the Internet with specific protocol versions (see Section \ref{subsub:nonhp} in the evaluation). Fingerprinting honeypots only with banner checks is prone to false positives, and therefore, we subject the instances to further protocol level and system-level checks. 

\paragraph{\textit{HTTP} Static Response}
The filtered instances with \textit{HTTP} and \textit{HTTPS} service identified in previous stages are checked for static \textit{HTTP} content in their response. Honeypots emulating the web services offer limited content by default which can be identified. The instances are queried with an \textit{HTTP GET} request to fetch the content and then match the static default content offered by the honeypots. Table \ref{tab:hp-http} in the Appendix shows the \textit{HTTP} response returned by honeypots. Upon match of static content, the instance continues to the next stage. 
This technique was adapted from \cite{Vetterl2018,morishita} for fingerprinting \textit{HTTP}-based honeypots. 

\paragraph{SSL/TLS Certificate check}
This stage compares certificate-specific attributes to known values from default certificates provided by honeypots. Some honeypots offer hard-coded \acrshort{tls} certificates that can be leveraged to fingerprint honeypot instances. Though there is a change of fingerprint on each certificate, attributes like issuer and provider remain static. For example, the Dionaea honeypot contains a certificate issued by a provider name that is consistent in all its deployments \cite{papazis2019detecting}. 
The \textit{SSL/TLS Certificate check} component stage checks the attributes \textit{certificate issuer} and the \textit{common subject name} of the certificate retrieved from web servers to identify Dionaea honeypots on the Internet.
Algorithm \ref{algo:cert-check}, in the Appendix, represents the pseudo-code block that checks an instance for Dionaea's default certificate parameters.

\paragraph{Protocol Handshake}
\label{subsub:ph}
The communication of systems over any network is established upon the negotiation of various communication parameters, before building a channel. Honeypots offer limited emulation and communication preferences. This limitation is caused due to the honeypot design or the utilization of certain protocol emulation libraries. 
We exploit this limitation of deviated behavior, in the protocol negotiation process, to identify honeypots. 
First, we observe the deviation in the negotiation process and the limited availability of parameters by establishing communication with in-house lab honeypots (see Section \ref{subsec:evalsetup}). 
We develop probes that attempt to establish a connection through limited parameters and observe the response for deviation for all emulated services.
Table  \ref{tab:hp-ph} summarizes the responses for certain negotiations of protocols.
We observe protocol handshake deviations that cause the acceptance of malformed request packets, return limited options for negotiation, or disconnect the session with an arbitrary message that is different from non-honeypot implementations. Algorithm \ref{algo:ph} in the Appendix describes the protocol handshake checks. The algorithm accepts a list of instances with their \acrshort{ip} address and port. For each instance, a request is sent for session initiation with specific parameters. The response is analyzed for deviations that match the response from honeypots. Upon match, the flag \textit{isDeviated} is set and such instances progress to the next framework stage. 

\begin{table*}[h]
\begin{tabular}{|l|l|l|l|}
\hline
\textbf{Honeypot} & \textbf{Protocol} & \multicolumn{1}{c|}{\textbf{Request}} & \multicolumn{1}{c|}{\textbf{Response}} \\ \hline
Kippo & SSH & \begin{tabular}[c]{@{}l@{}}SSH-2.0-OpenSSH \textbackslash{}n\textbackslash{}n\textbackslash{}n\textbackslash{}n\textbackslash{}n\textbackslash{}n\textbackslash{}n\textbackslash{}n\textbackslash{}n\textbackslash{}n\end{tabular} & \begin{tabular}[c]{@{}l@{}}"bad packet length *" or\\ "protocol mismatch\textbackslash{}n"\end{tabular} \\ \hline
Cowrie & SSH & \begin{tabular}[c]{@{}l@{}}1. SSH-2.0-OpenSSH\_6.0p1 Debian-4+deb7u2 \textbackslash{}n\\ 2. SSH-2.0-OpenSSH\_6.0p1 Debian-4+deb7u2 \textbackslash{}n\end{tabular} & "protocol mismatch\textbackslash{}n" \\ \hline
Gaspot & Telnet & I30100 & 9999FF1B \\ \hline
Conpot & S7 & \begin{tabular}[c]{@{}l@{}} "H", "0300002102f0803207000000000008
\textbackslash{}n\\00080001120411440100ff09000400110001"\textbackslash{}n\end{tabular} & 0x32 \\ \hline
Conpot & Modbus & \begin{tabular}[c]{@{}l@{}} function\_code': None, 'slave\_id': 0, \textbackslash{}n\\ 'request': '000000000005002b0e0200'  \textbackslash{}n\end{tabular} & Disconnection  \\ \hline
Glastopf & HTTP & GET /HTTP/1.0 & Server: BaseHTTP/0.3 Python/2.5.1 \\ \hline
Dionaea & HTTP & GET /HTTP/1.0 & Server: nginx \\ \hline
Amun & HTTP & GET HTTP/1.1 & Server: Apache/1.3.29 \\ \hline
MTPot & Telnet & WILL (251) Linemode & Won't (252) Linemode \\ \hline
\end{tabular}
\caption{Protocol handshake deviation}
\label{tab:hp-ph}
\end{table*}

\paragraph{Library Dependency Check}
Emulations in low and medium-interaction honeypots are often developed by referring to external libraries. Libraries offer limited emulation capabilities based on their design and frequently return static values in certain queries. Furthermore, some libraries referred by honeypots have not been well maintained. Vetterl et al. have leveraged the use of libraries in honeypots to craft specific probes that return static values~\cite{Vetterl2018}. This static information can be used to fingerprint the honeypots. Table \ref{tab:library} shows the libraries used by many well-known honeypots for the service emulation and their last update. 
Leveraging the aforementioned static implementation and limited emulation, we develop 
the probes based on \cite{Vetterl2018} that request for specific information from the end-systems. We compare the response to known static responses from the honeypots. We proceed in case of a match. 
In honeypots, protocol handshake is also dependent on the library used for emulation purposes and hence these two stages are intertwined. Nevertheless, we use this check to check for additional dependencies that can signal static behavior.

\begin{table}[]
\centering
\begin{tabular}{|l|l|l|l|}
\hline
\multicolumn{1}{|c|}{\textbf{Honeypot}} & \multicolumn{1}{c|}{\textbf{Protocol}} & \multicolumn{1}{c|}{\textbf{Library}} & \multicolumn{1}{c|}{\textbf{Updated}} \\ \hline
Kippo & SSH & TwistedConch & May2015 \\ \hline
Cowrie & SSH & TwistedConch & May2018 \\ \hline
MTPot & Telnet & telnetsrv & Dec2012 \\ \hline
Cowrie & Telnet & TwistedConch & May2018 \\ \hline
Dionaea & HTTP & custom & Sep2016 \\ \hline
Glastopf & HTTP & BaseHTTPServer & Oct2016 \\ \hline
Conpot & HTTP & BaseHTTPServer & Mar2018 \\ \hline
\end{tabular}
\caption{Library references in honeypots}
\label{tab:library}
\end{table}

\paragraph{Static Command Response}
\label{subsub: scr}
Due to the nature of honeypots, developers are compelled to implement some services with static responses or disconnect the communication for specific command requests. For instance, some honeypots attempt to overcome such issues via a static response (e.g. \textit{"Invalid Command"}), or disconnect with the user. 
We leverage this gap in implementation for having probes request systems with commands to expect known static responses from the end systems. Table \ref{tab:hp-sr} shows the static response returned by honeypots for specific commands by our probes.

\subsubsection{Metascan-based Fingerprinting Pipeline}
Metascan-based techniques aim at honeypot detection using passive-fingerprinting techniques. Our framework uses information available through Shodan and Censys to determine if an instance is a honeypot. The metascan-pipeline consists of \textit{four} stages based on the underlying protocol. Although some SotA, e.g. \cite{morishita}, have used mass-scan engines to search for honeypot signatures, we use persistent checks in our stages to assure that the instance is a honeypot. We use checks to determine if the network belongs to a research facility, has an identified domain attached to it, or if the instance is on a cloud infrastructure. This information helps to further distinguish the honeypots by analyzing operational parameters. 

\paragraph{Shodan and Censys Search}
Contrary to the probe-based scanning that requires us to use a tool to perform the scan, we leverage the available data from Shodan and Censys that perform the scans daily. We search the platforms for systems with open ports concerning the services emulated by honeypots in our tests. The result of the search provides a list of instances that undergo further fingerprinting process. Both Shodan and Censys provide \acrshort{api}s for querying their databases. Algorithm \ref{algo: metsearch} in the Appendix shows the procedure for the search performed on Shodan and Censys. The search results return an \acrshort{ip} address and port, for the identified instances. 

\paragraph{Keyword Search}
\label{subsub: keyword}
Shodan and Censys store information about the systems exposed to the Internet that include banners, web content, protocol negotiation parameters, and more. In addition to system-specific information, they also provide metadata about the \acrshort{ip} address allocated to the system like geo-location, \acrshort{isp}/\acrshort{as} and the hosting provider. The degree of information and the format available on these databases vary based on the techniques followed by the mass-scan engine. We leverage such information to filter the instances obtained in the previous step. The search is performed with keywords identified from the probe-based stages like static content, banners, and protocol negotiations. Table \ref{tab:hp-ks} shows the used keyword parameters for filtering instances in Shodan and Censys. The resulting data contains a list of instances of systems with specific ports and matching filtered criteria.

\paragraph{\acrshort{isp} and \acrshort{as} Check}
Honeypots are also classified based on their usage in research and production environments. Research organizations deploy honeypots to gather attack-data for threat intelligence research. Enterprise systems deploy honeypots for proactive attack detection. Following the previous stages, we examine whether the instance is part of a research organization or an institute. It is also possible that an enterprise company may be hosting a production honeypot with an unassigned domain. For instance, the honeypots deployed in our lab lie under the university \acrshort{as} while they do not have a domain registered to them. To cope with this, this component checks the WHOIS database to search for information about the network to which the system is attached to.

\paragraph{Cloud Hosting Check}
Cloud infrastructure enables defenders to set up and deploy honeypots on cloud environments to easily gather attack-data. Many honeypot developers offer a container-based configuration of honeypots for easy installation and deployment. As a result, many honeypot instances can be found in cloud instances. 
We argue that many honeypots are deployed in cloud environments though they are logically invalid for the emulated infrastructure. For example, we find many instances of Conpot, an \acrshort{ics} based honeypot, which emulates industrial cyber-physical systems. However, it is improbable to find \acrshort{ics} devices on cloud networks. This component checks whether instances related to specific \acrshort{ics} protocols (i.e. Modbus and S7) are deployed on a cloud infrastructure. 

\subsubsection{Fully Qualified Domain Name (FQDN) Check}
A \acrshort{fqdn} is allocated to an Internet-facing system to avoid memorization of the \acrshort{ip} addresses. We perform a check to examine whether the identified instances from both pipelines have an assigned \acrfull{dns} domain. Honeypot systems, by design, are fake systems and are unlikely to have domain names allocated as it is risky for the organizations deploying them. For instance, an attacker may claim to have found a vulnerable or compromised system belonging to an enterprise domain, resulting in negative publicity for an organization. 
Therefore, administrators, in principle, avoid assigning a domain/\acrshort{dns} for the honeypots. We utilize this understanding of the administrators and filter the \acrshort{ip} addresses received from the \acrshort{ip} pool to find systems without domain names assigned. The \acrshort{ip} addresses that do not have a \acrshort{dns} are transitioned to the next state. 
The FQDN check differs from the \acrshort{as} check, in a way that it checks for any domain associated with the IP address, while the AS check performs a lookup of the IP address allocation by the \acrshort{as} to an entity. The information about the AS and the ISP helps in identifying the type of entity, for example, a research organization or honeypot instance in a production network of an organization. 

\subsubsection{Framework Output}
The output state of the framework provides a list of instances that are inferred as honeypots from our fingerprinting framework. The list contains instances from both the probe-based and the metascan-based honeypots.    

\section{Evaluation}
\label{sec:eval}

We evaluate the ability of the proposed multistage honeypot fingerprinting framework in discovering honeypots. The evaluation considers \textit{nine} honeypot implementations and specifically focuses on \textit{nine} protocols as listed in Table \ref{tab:honeypots}. 
The choice of honeypots is based on a number of factors. First, these honeypots are considered some of the most popular ones and most frequently deployed (see e.g. the ENISA recommendations in \cite{grudziecki2012proactive}). Moreover, these represent the honeypots examined in the majority of the related work (cf. Section \ref{sec:rw}), which provides us the ability to make some comparisons (e.g. with \cite{Vetterl2018,morishita,zamiri2019gas}). Lastly, all of the selected honeypots are open-source implementations. 

Our main goal is to examine how many honeypots the framework can identify. We highlight here that the absence of ground truth data for honeypots is a known problem in the field. However, we argue that the multistage nature of the framework highly reduces the probability for false positives (we further discuss this issue in Section \ref{subsec:groundtruth}). 
In addition, we want to determine the relation between the probe-based and metascan-based detection. Our hypothesis is that the probe-based pipeline should produce significantly better results. Still, the question of whether the metascan pipeline can identify honeypots beyond the ones already identified via the probe-based methods is an open question that we will attempt to answer. 
Lastly, we are interested in further examining encounters with IP addresses that pass some, but not all, of our tests. We believe that these systems might be vulnerable ones, which can easily be exploited by adversaries.

\subsection{Lab environment tests}
First, we deploy all the honeypot implementations (see Table \ref{tab:honeypots}) in a lab environment and test all probes which are implemented to collect state-specific information like banners, static content, protocol handshake and static command responses. We confirm that honeypots test positive for \textit{all} the different modules (see Figure \ref{fig:fsm}) of the probe-based phase. Following these tests, we evaluate the multistage framework against the known honeypot instances in the lab environment. All the honeypot instances were successfully detected by our framework. 

\begin{table}[h!]
\centering
{%
\begin{tabular}{|c|l|c|}
\hline
\textbf{Honeypots} & \multicolumn{1}{c|}{\textbf{Ports \& Services}} & \textbf{Version} \\ \hline
Kippo & \begin{tabular}[c]{@{}l@{}}Ports:22/2222\\ Services: SSH\end{tabular} & 0.9 \\ \hline
Cowrie & \begin{tabular}[c]{@{}l@{}}Ports: 22/2222 23/2323\\ Services: SSH, Telnet\end{tabular} & 2.1.0 \\ \hline
Glastopf & \begin{tabular}[c]{@{}l@{}}Ports: 80, 8080\\ Services: HTTP\end{tabular} & 3.1.2 \\ \hline
Dionaea & \begin{tabular}[c]{@{}l@{}}Ports: 80, 443, 21\\ Services: HTTP, FTP\end{tabular} & 0.9.0 \\ \hline
Nepenthes & \begin{tabular}[c]{@{}l@{}}Ports: 21\\ Services: FTP\end{tabular} & 0.2.2 \\ \hline
Amun & \begin{tabular}[c]{@{}l@{}}Ports: 23,21,80,36,143\\ Services: Telnet, FTP, HTTP, SMTP, IMAP\end{tabular} & 0.2.3 \\ \hline
Conpot & \begin{tabular}[c]{@{}l@{}}Ports: 80, 502, 102\\ Services: HTTP, Modbus, S7\end{tabular} & 0.5.2 \\ \hline
Gaspot & \begin{tabular}[c]{@{}l@{}}Ports: 100001\\ Services: ATG\end{tabular} & base \cite{wilhoitgaspot} \\ \hline
MTPot & \begin{tabular}[c]{@{}l@{}}Ports: 23\\ Services: Telnet\end{tabular} & base \cite{MTPot}\\ \hline
\end{tabular}%
}
\caption{Honeypots tested in our internal lab environment}
\label{tab:honeypots}
\end{table}

\subsection{Evaluation Setup}
\label{subsec:evalsetup}
After performing the aforesaid experiments, we are now ready to perform an Internet-wide scan. 
We use the ZMap tool as our scanning tool \cite{zmap} to scan a total of $2.9$ billion IP addresses\footnote{ZMap excludes a number of IP addresses from its scan by default; these include reserved and unallocated IP space.}. Our tests follow the flow of Figure \ref{fig:fsm}. That is, we first perform a probe-based scan and afterwards perform an independent metascan by making use of Shodan and Censys \cite{SHODAN,censys}.  
Our experiments were conducted in a period of six months. The experiment is carried out as $3$ scanning periods, for the entire-framework. 
The metascan-based approach was relatively faster to perform the search and analysis, although Shodan and Censys enforce rate limiting on the API requests. Over a period of six months, we conducted three iterations. The results depicted in the following sections provide a summation of all the unique honeypot instances identified from the three scan iterations.

We, once more, highlight that this article does not take into account high interaction honeypots. This is due to the very different characteristics of high interaction honeypots (i.e. real systems instead of emulated ones); in fact, this is the case with all the SotA (e.g. \cite{Vetterl2018,morishita,zamiri2019gas}). Hence, both our article as well as all existing related work are prone to false negatives.

\subsection{Results}
By firstly performing a ZMap scan, we derive Table \ref{tab:system-protocols} that shows the number of identified systems (not necessarily honeypots) on the Internet that exhibit relevant open ports. Subsequently, the framework performs the various checks shown in Figure \ref{fig:fsm}.

\begin{table}[h!]
\centering
\begin{tabular}{|c|c|c|}
\hline
\textbf{Protocol} & \multicolumn{1}{l|}{\textbf{Port}} & \multicolumn{1}{l|}{\textbf{\begin{tabular}[c]{@{}l@{}} No. of Systems on the \\ Internet (in Million)\end{tabular}}} \\ \hline
HTTP & \begin{tabular}[c]{@{}c@{}}80,8080,\\ 8888\end{tabular} & 67.31 \\ \hline
HTTPS & 443 & 56.06 \\ \hline
SSH & 22 & 18.65 \\ \hline
FTP & 21 & 10.39 \\ \hline
SMTP & 25 & 7.71 \\ \hline
Telnet & 23 & 5.27 \\ \hline
\end{tabular}
\caption{Number of identified instances and protocols/ports}
\label{tab:system-protocols}
\end{table}

\subsubsection{Honeypot identification}
\begin{figure}[h!]
    \centering
    \includegraphics[width=0.85\columnwidth]{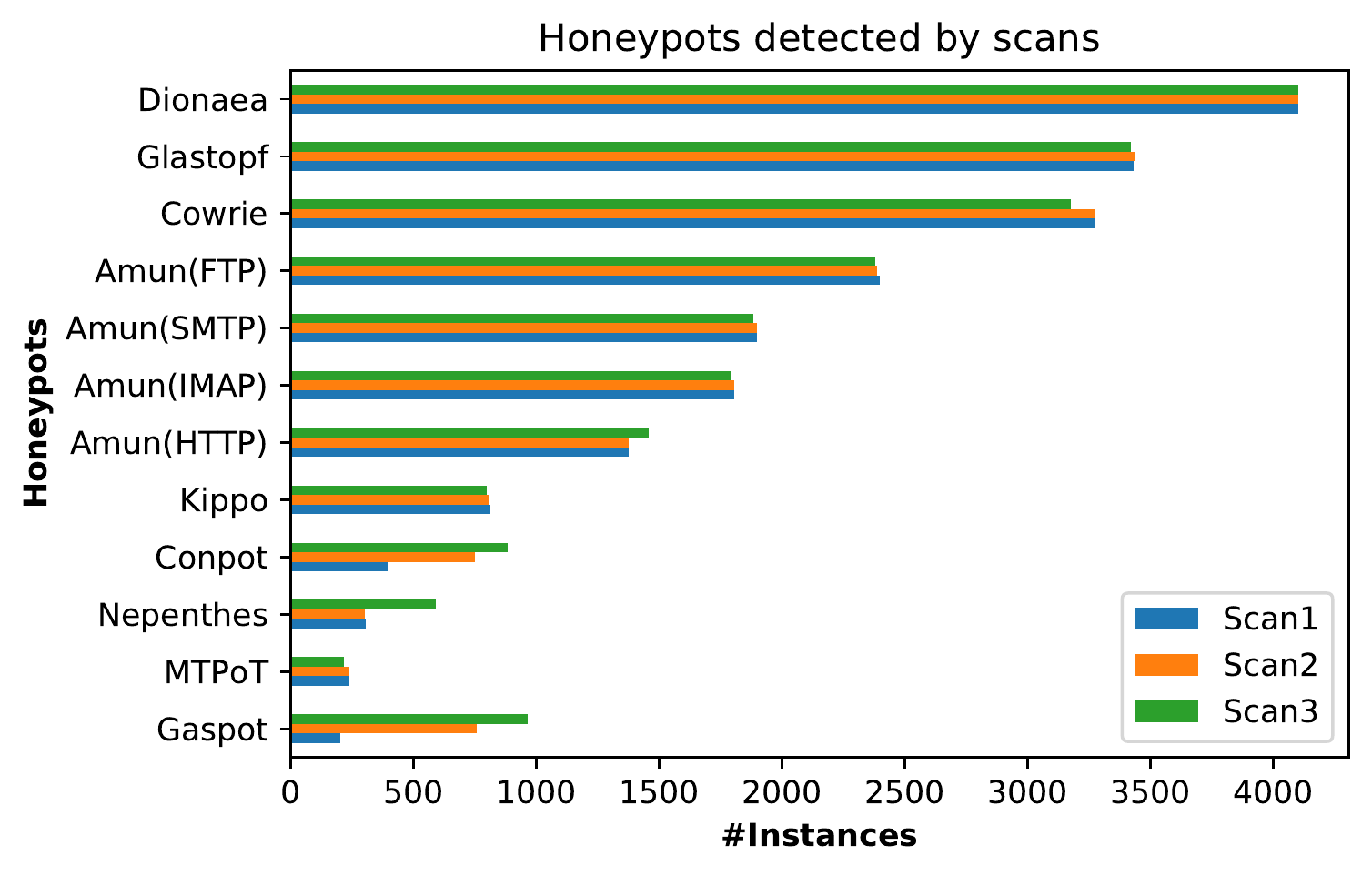}
    \caption{Honeypots detected per scan}
    \label{fig:scans}
\end{figure}

\begin{figure}[h!]
    \centering
    \includegraphics[width=0.85\columnwidth]{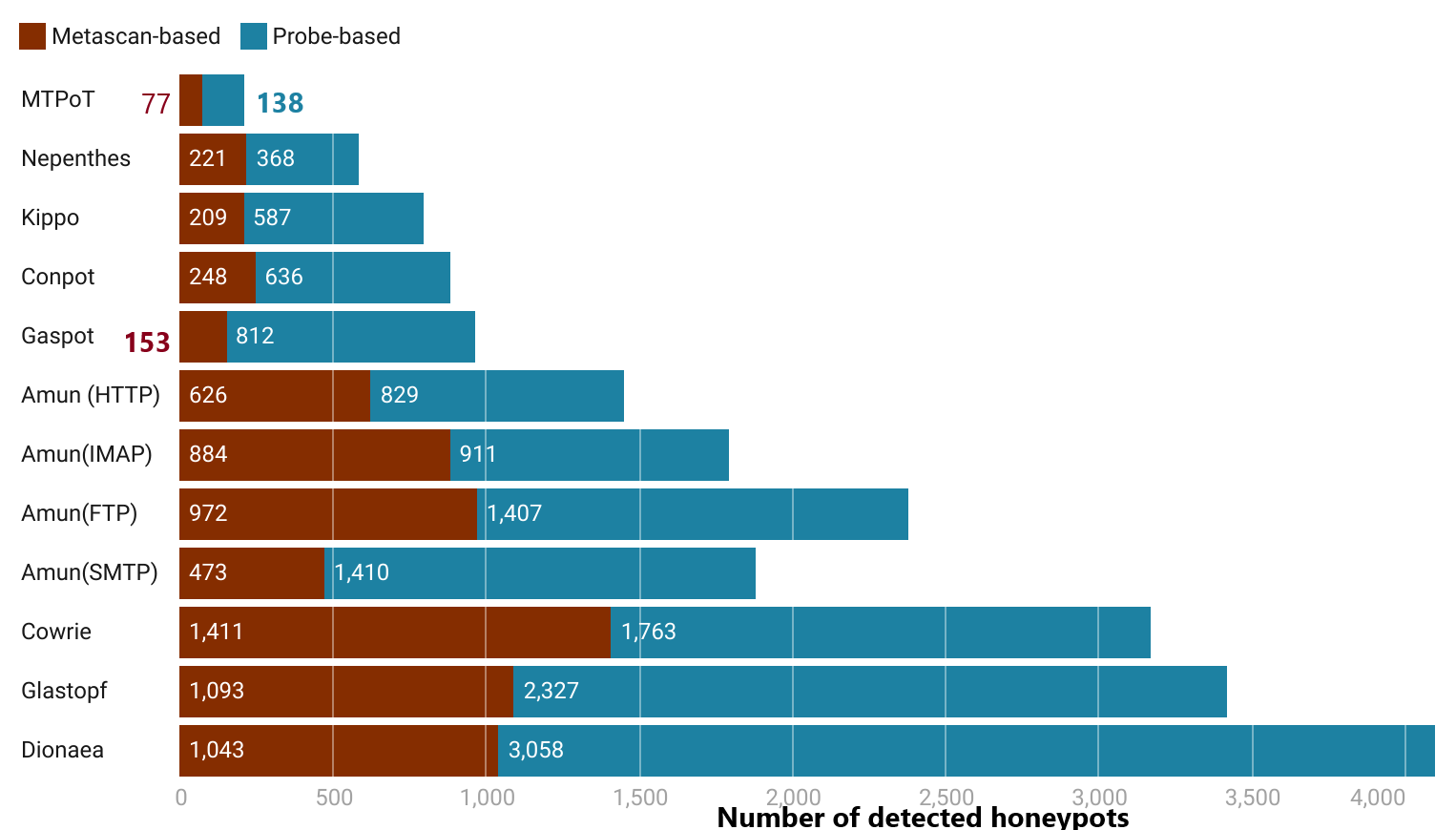}
    \caption{Honeypots detected by type and technique}
    \label{fig:results}
\end{figure}

Overall, the framework detected a total of $21,855$ honeypots. Figure \ref{fig:scans} shows the honeypot instances detected over three sequential scans over a period of $6$ months. 
Figure \ref{fig:scans} also depicts the change in honeypot instances detected over the $3$ scans. The instances of honeypots Gaspot, Conpot and Amun (HTTP) were detected more in the third scan while the others remained constant or reduced. This could be because of honeypots instances undergoing either a churn or because they were simply blocked/offline. We discuss this further in Section \ref{subsec:groundtruth}.  The metascan-based technique has identified $7,410$ unique honeypots and the remaining $14,246$ were detected by the probe-based technique. Figure \ref{fig:results} summarizes the honeypots detected by probe-based and metascan-based approaches for each honeypot. The numbers on the bars indicate the \textit{unique} instances detected by the approaches and scans.  
An interesting finding is that \textit{all IP addresses identified as honeypots by the metascan-based approach were already detected by the probe-based approach}. This is important as it confirms our hypothesis that probe-based is superior to the metascan. In fact, this suggests that the metascan pipeline can be ignored without any loss of information.

\begin{figure}[h!]
    \centering
    \includegraphics[width=0.85\columnwidth]{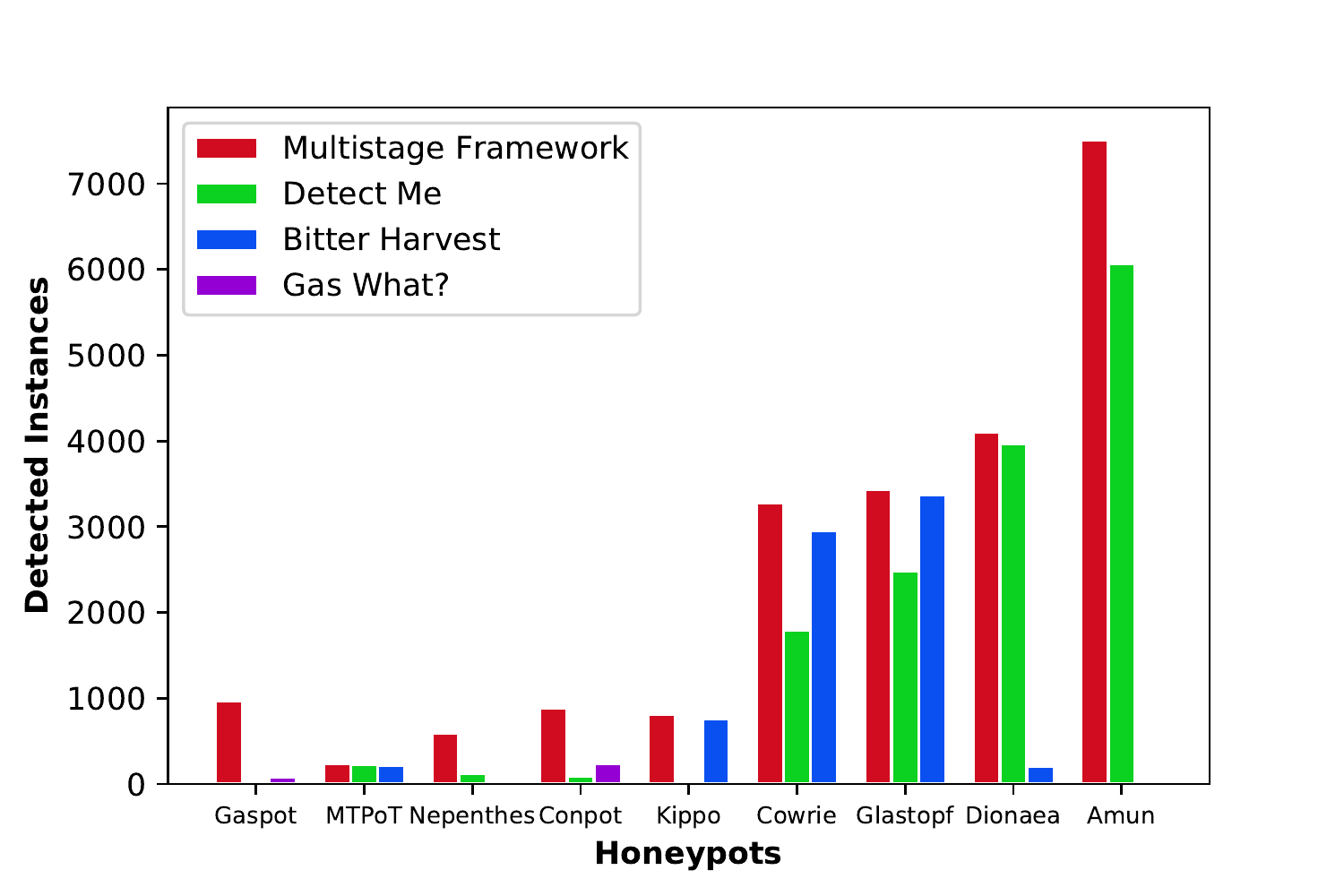}
    \caption{Comparison to previous measurements in related work}
    \label{fig:sota}
\end{figure}

Figure \ref{fig:sota}, compares our findings with the SotA measurements from Vetterl et al. (\textit{Bitter Harvest}) \cite{Vetterl2018}, Morishita et al. (\textit{Detect Me}) \cite{morishita} and Zamiri et al. (\textit{Gas What?})  \cite{zamiri2019gas}. The figure shows the total honeypot instances detected by SotA in comparison to our approach. We note that the honeypots Nepenthes and Amun were not evaluated by \cite{Vetterl2018}; in addition, \cite{zamiri2019gas} only evaluated Gaspot and Conpot honeypots.
We want to highlight that the value of this figure does not lie within the improved results on the majority of the honeypots.
Direct comparison with previous measurements is not adequate due to the different time frame.
Instead, we argue that these results suggest a number of interesting findings. First, they independently confirm previous studies' conclusions with regard to the global (poor) state of honeypot deployments \cite{Vetterl2018}. Second, our results come more than one year after the aforesaid studies: this provided a relatively long period for honeypot administrators to react, while many honeypots (e.g. Conpot) have been updated to fix relevant vulnerabilities. Lastly, the multistage nature of our framework suggests that, in contrast to related work, we should encounter a very small number of false positives. That is, IP addresses are only marked as honeypots when all (relevant) stages are confirmed.

We emphasize once more that our results are not directly comparable with the SotA as they come in a different time-period from the reference work. This suggests, that honeypot administrators had time to apply updates fixing fingerprinting issues, many honeypots might have been deactivated over time and others might have been introduced. That said, our results indicate that many honeypot administrators are not properly managing their systems.

\subsubsection{Honeypot Versions}
We determine the versions of the instances detected as honeypots by examining specific changes added to the honeypots through patches released by the developers. However, versions could not be determined for some honeypots that do not maintain releases (i.e. MTPot and Gaspot). We find that the majority of the honeypots detected, have not been updated by the administrators even though there were patches released by the honeypot developers (e.g. for certain fingerprinting attacks). Furthermore, we detect instances running on honeypots that are no longer maintained by the developers. The developers of these honeypots disclose that the project has been discontinued and also suggest newer honeypots under active maintenance. We list the instances with the identified deployed versions in Table \ref{tab:hp-version}.

\begin{table}[]
\centering
\begin{tabular}{|c|c|l|}
\hline
\textbf{Honeypot}                                         & \textbf{\begin{tabular}[c]{@{}c@{}}Deployed\\  Version\end{tabular}} & \multicolumn{1}{c|}{\textbf{\#Instances}}               \\ \hline
\begin{tabular}[c]{@{}c@{}}Conpot\\ 0.5.2*\end{tabular}   & \begin{tabular}[c]{@{}c@{}}0.5.2\\  0.5.0\\  0.4.0\end{tabular}      & \begin{tabular}[c]{@{}l@{}}221\\ 496\\ 167\end{tabular} \\ \hline
\begin{tabular}[c]{@{}c@{}}Cowrie\\ 2.1.0*\end{tabular}   & \begin{tabular}[c]{@{}c@{}}2.1.0\\  1.5.3\\  1.5.1\end{tabular}      & \begin{tabular}[c]{@{}l@{}}17\\ 232\\ 2925\end{tabular} \\ \hline
\begin{tabular}[c]{@{}c@{}}Glastopf\\ 3.1.2*\end{tabular} & \begin{tabular}[c]{@{}c@{}}3.1.2\\  0.2.0\end{tabular}               & \begin{tabular}[c]{@{}l@{}}4\\ 3416\end{tabular}        \\ \hline
\begin{tabular}[c]{@{}c@{}}Dionaea\\ 0.8.0*\end{tabular}  & \begin{tabular}[c]{@{}c@{}}0.8.0\\  0.6.0\end{tabular}               & \begin{tabular}[c]{@{}l@{}}2259\\ 1782\end{tabular}     \\ \hline
\end{tabular}
\caption{Detected honeypot versions (* latest version)}
\label{tab:hp-version}
\end{table}

\subsubsection{Honeypots with Default Configuration}
The honeypots considered in our tests can be deployed with a default configuration. Nevertheless, for some honeypots, the developers explicitly provide additional templates and guidelines to change the default settings. The usage of default honeypot configuration can be problematic as it makes fingerprinting significantly easier.

To determine this, we compare the cumulative results from the framework's \textit{HTTP Static Response} and the \textit{Static Command Response} stages to the default configuration of the deployed honeypots in our lab environment (see Section \ref{subsec:evalsetup}).
Therefore, upon matching, we can infer that the instance is a honeypot deployed with its default configuration. 
We observe that the majority of the detected honeypots are running with default configurations that make primitive fingerprinting techniques like static http-content very successful. 
We list the number of honeypot instances running with default configurations in Table \ref{tab:hp-dc}.

\begin{table}[h!]
\centering
\begin{tabular}{|c|c|c|}
\hline
\textbf{Honeypots} & \textbf{\begin{tabular}[c]{@{}c@{}}\#Instances with\\ default configuration\end{tabular}} & \textbf{\begin{tabular}[c]{@{}c@{}}\#Instances without\\ default configuration\end{tabular}} \\ \hline
Gaspot & 925 & 40 \\ \hline
MTPot & 215 & 0 \\ \hline
Conpot & 777 & 107 \\ \hline
Nepenthes & 531 & 58 \\ \hline
Kippo & 773 & 23 \\ \hline
Cowrie & 3149 & 25 \\ \hline
Amun & 7455 & 57 \\ \hline
Glastopf & 3416 & 4 \\ \hline
Dionaea & 4064 & 37 \\ \hline
\textbf{Total} & \textbf{21305} & \textbf{351} \\ \hline
\end{tabular}
\caption{Detected honeypots running on default configuration}
\label{tab:hp-dc}
\end{table}

\subsubsection{Non-honeypot encounters}
\label{subsub:nonhp}
As a result of multistage checks from the framework, instances are filtered out at each stage when they fail the matching criteria. 
We further analyze the non-honeypot instances that were filtered out at multiple stages to determine the cause of filtration at a particular stage and/or the success in other stages. 
Table \ref{tab:faile-hp} in the Appendix shows the non-honeypot instances determined at stages in our framework based on honeypot types. Based on this, we derive the following findings.
\begin{table}[h!]
\centering
\begin{tabular}{|c|l|}
\hline
\textbf{Vulnerability} & \multicolumn{1}{c|}{\textbf{\# Instances}} \\ \hline
\multicolumn{2}{|c|}{\textbf{Default passwords (SSH)}} \\ \hline
\multicolumn{1}{|l|}{\begin{tabular}[c]{@{}l@{}}root, root\\ admin, admin\\ root, 1234\\ admin, 1234\\ root, 123456\\ root, (no password)\\ admin, (no password)\end{tabular}} & \begin{tabular}[c]{@{}l@{}}216\\ 124\\ 23\\ 43\\ 21\\ 18\\ 28\end{tabular} \\ \hline
\multicolumn{2}{|c|}{\textbf{Default passwords (FTP)}} \\ \hline
\multicolumn{1}{|l|}{\begin{tabular}[c]{@{}l@{}}root, root\\ admin, admin\\ root, 1234\\ admin, 1234\end{tabular}} & \begin{tabular}[c]{@{}l@{}}94\\ 29\\ 19\\ 8\end{tabular} \\ \hline
\multicolumn{2}{|c|}{\textbf{Vulnerable Banners (SSH)}} \\ \hline
\multicolumn{1}{|l|}{\begin{tabular}[c]{@{}l@{}}SSH-2.0-ROSSSH\\ SSH-2-0-libssh-0.7.0(5)\end{tabular}} & \begin{tabular}[c]{@{}l@{}}263,516\\ 196\end{tabular} \\ \hline
\multicolumn{2}{|c|}{\textbf{Vulnerable Banners (FTP)}} \\ \hline
\multicolumn{1}{|l|}{\begin{tabular}[c]{@{}l@{}}220 ProFTPD 1.3.5 Server\\ 220 ProFTPD 1.3.1 Server\\ 220 Serv-U FTP Server v6.2\end{tabular}} & \begin{tabular}[c]{@{}l@{}}53,873\\ 15,823\\ 21,023\end{tabular} \\ \hline
\end{tabular}
\caption{Vulnerable non-honeypot instances}
\label{tab:vul-nhp}
\end{table}

\paragraph{SSH and FTP Instances with Default Passwords} We find SSH instances running on default passwords that met the initial criteria for SSH honeypot detection in our framework, but fail in other stages (e.g. static command and library checks). These instances' credentials match the ones of the default passwords accepted by Kippo and Cowrie honeypots. Our conclusion is that these are either vulnerable devices with default logins or high-interaction honeypots. We list the number of vulnerable SSH instances found with default passwords in Table \ref{tab:vul-nhp}.

\paragraph{SSH and FTP Instances with Vulnerable Versions}
From the instances that were filtered out of the banner check stage (in the probe-based pipeline), we identify the number of instances that appear to contain vulnerable versions in their banners. In particular, we take into account banners that have a high severity vulnerability (by making use of the National Vulnerability Database \cite{booth2013national}). 
We identify a total of $263,712$ instances with vulnerable versions as per the advertised banners. The banners and the number of instances identified are listed in Table \ref{tab:vul-nhp}.

\subsubsection{Experiment repetition: gain and blocked/offline instances}
Due to the nature of our experiments (i.e. long time windows and rather aggressive fingerprinting scans) we expect that:\textit{ i)} we will observe some fluctuation in our results, \textit{ii)} we will have some gain as new honeypots are introduced on the Internet, \textit{iii)} we expect some of the networks to blocklist our scanners, and lastly \textit{iv)} we anticipate some honeypots not to be responsive due to them taken down, maintenance and/or network errors.

We scan the Internet with a different scanning host that has different \acrshort{ip} address and subnet. We compare the results from the different scanning periods to identify new and existing honeypot instances. In the next step, we analyze the \acrshort{ip} address of the new honeypot instances detected against our framework and check the \acrshort{ip} address for their subnet and their \acrshort{as}. If the \acrshort{ip} address belongs to a different subnet but belongs to the same \acrshort{as}, and further matches to the properties of the honeypot identified in the previous period, we infer that the honeypots are the same but had some churn-related effects. Moreover, we further examine the gain vs. blocked trade-off by trying to connect to the new \acrshort{ip} address of the honeypot instance from our previous scanning host. If the honeypot instance blocks the connection from the first connected host but was connected by the second scanning host, then it is very likely that the honeypot administrator has blocklisted the \acrshort{ip} address of the first scanning host.

Table \ref{tab:churn} shows the number of new honeypot instances detected in the scans and the instances that were either blocked or offline. 
There was a significant number of new Nepenthes honeypots instances detected in the third scan. On tracing the \acrshort{ip} addresses of the new instances, we find that all the new detected honeypots were hosted by a hosting provider which was traced earlier hosting Nepenthes instances on another subnet. We can infer that either the honeypots were configured to undergo some \acrshort{ip} rotation logic or were simply offline for a certain period. Overall, we find that only $2.3$\% of the honeypot instances have changed their \acrshort{ip} and only $1$\% are not offline after the first scan. 

\begin{table}[h!]
\centering
\begin{tabular}{|c|c|c|c|c|}
\hline
\textbf{Honeypot} & \textbf{\begin{tabular}[c]{@{}c@{}}Scan-2\\ New Instances\end{tabular}} & \textbf{\begin{tabular}[c]{@{}c@{}}Scan-2\\ Blocked/Offline\end{tabular}} & \textbf{\begin{tabular}[c]{@{}c@{}}Scan-3\\ New Instances\end{tabular}} & \textbf{\begin{tabular}[c]{@{}c@{}}Scan-3\\ Blocked/Offline\end{tabular}} \\ \hline
Gaspot & 567 & 12 & 387 & 11 \\ \hline
MTPoT & 0 & 1 & 0 & 23 \\ \hline
Nepenthes & 0 & 3 & 573 & 16 \\ \hline
Conpot & 367 & 33 & 110 & 23 \\ \hline
Kippo & 0 & 4 & 0 & 13 \\ \hline
Amun & 0 & 3 & 63 & 51 \\ \hline
Cowrie & 0 & 4 & 0 & 98 \\ \hline
Glastopf & 3 & 2 & 0 & 13 \\ \hline
Dionaea & 0 & 0 & 0 & 0 \\ \hline
\end{tabular}
\caption{Identification gain vs. blocked/offline instances}
\label{tab:churn}
\end{table}

\subsection{Ground Truth Validation}
\label{subsec:groundtruth}
The absence of ground truth knowledge regarding honeypots creates a challenging landscape for measuring metrics such as precision or possible false positives. This is a fundamental problem in the area of honeypot fingerprinting that cannot be solved in its entirety. Hence, in the following we attempt to provide indications on why false positives are not a significant issue in our approach.

First, in contrast to the state of the art, we propose a framework that requires multiple steps to be confirmed until an IP address is marked as a honeypot. These steps include a multitude of independent checks which, we argue, significantly decrease the probability of false positives. Looking at the SotA, Vetter et al. \cite{Vetterl2018} measure the detection accuracy  using the responses received from the honeypots by generating a cosine similarity score and Morishita et al. \cite{morishita} use the matching of honeypot signatures in four datasets. In contrast, our approach relies on multiple checks at each stage to minimize false positives.

Second, we replicate and extend the ground truth validation proposed by \cite{morishita} and \cite{vogt2007army}. Morishita et al. argue that a honeypot IP address cannot be present in IP spaces that are known for their commercial usage. This argument obviously does not solve the absence of ground truth, but rather provides a minor indication that the identified IP addresses are not clear false positives. Vogt et al. also use a similar validation in their evaluation to check if the domain identified by their model is listed on sources providing web-statistics like the top $1$ million domains \cite{vogt2007army}. 
In this context, we evaluate our results by comparing the identified honeypot IP addresses with the top $1$ million domains from Alexa \cite{Alexa}, Majestic \cite{Majestic} and Cisco-Umbrella \cite{Cisco-Umbrella} with known benign FTP servers, as well as known university SMTP domains. 
For this evaluation we fetch the Alexa top $1$ million domains from Alexa, perform a \acrshort{dns} lookup and examine whether our results match them. Similarly we fetch the top $1$ million domains from the Majestic and the Cisco-Umbrella websites. We confirm that none of \acrshort{ip} address from these domains are found in our results. We note that the \acrshort{ip} addresses for some of the domains change based on the geo-location of resolution due to the \acrfull{cdn}. Hence, we repeated the experiments by connecting to many different geo-locations by using a \acrfull{vpn} provider.
Moreover, we fetch the list of official FTP mirrors from GNU \cite{gnu}, Apache \cite{Apache}, Ubuntu \cite{Ubuntu}, Debian  \cite{Debian} and find $1,231$ unique domain names. Upon performing a \acrshort{dns} lookup, we get $2,784$ \acrshort{ip} addresses. Once more, none of the identified honeypots match these IP addresses. Furthermore, we retrieve the list of university domain names from \cite{University} for evaluating Amun (SMTP) honeypot. Upon performing a \acrshort{dns} lookup we find $12,012$ \acrshort{ip} addresses. There were no honeypots detected in the domains from this list. 
To sum up, while the SotA uses a singular method to deal with false positives, our approach utilizes multiple stages. Moreover, we further test our results with an adaption and extension of the techniques employed by \cite{morishita} to address the absence of ground truth knowledge.

\section{Discussion}
\label{sec:discussion}
The evaluation of the multistage framework involved an experimental setup to reduce false positives and help in classification of honeypot instances. In this section we discuss the ethical considerations and experimental setup considered during the experimentation phase.

\subsection{Ethical considerations}
This section takes into account the various ethical considerations we had during our research.

\subsubsection{Experiments}
First, we inform the IT administrators of our organization about the ongoing research and seek their assistance for providing an approved setup for scanning the Internet. This is important as organizations tend to blocklist the \acrshort{ip} addresses of sources that appear to be scanning them. Second, we setup a website on the IP address of our scanner that provides a disclosure/explanation of our research purpose. This assists in limiting the effects of blocklisting the IP addresses of our organization.

\subsubsection{Results Disclosure}
The list of honeypot instances obtained through our framework is not publicly shared. We only present here aggregated statistics and do not share any identifiers of the honeypot instances. We seek guidance from the privacy department of our organization for guidelines on storing the results of our experiments and being compliant to GDPR. We followed the GDPR compliance by anonymizing the \acrshort{ip} addresses after three months following the completion of  our research.  

\subsubsection{Ethical disclosure: notifying Honeypot Developers}
We contact the honeypot developers of all of the active honeypot implementations and provide them with the specifics of the honeypot fingerprinting methods that can be used against them.  Moreover, we contacted members of the Honeynet project \cite{Honeynet}, an international security research organization that focuses on honeypot research, to further disclose the fingerprinting mechanisms that we have identified.

\subsubsection{Ethical disclosure: notifying Honeypot Administrators}
We take all $21,855$ IP addresses that were identified as honeypots and perform a WHOIS scan to find relevant contact information.
Based on this, we identify $939$ email addresses that correspond to all the IP addresses that we managed to find information about. We note that in many cases one email address corresponds to hundreds of honeypot instances, deployed in the same network.
There are multiple benefits from this procedure. First and foremost, we notify honeypot administrators that their deployments are vulnerable to our fingerprinting methods. Second, we ask administrators to contact us in case they are confident that our finding is a false positive and no honeypot deployment has taken place in their networks. This acts as an additional false positive sanity check. Until the time of submission, we did not receive any false positive claim from the contacted administrators. 

\subsection{Countermeasures against fingerprinting}
This section discusses potential countermeasures against fingerprinting. First, we want to emphasize that, due to their nature, low and medium interaction honeypots can always be identified upon continuous interaction and response analysis. Instead, we argue that the emphasis should be to reduce as much as possible fingerprinting vulnerabilities that can easily be automated.

\textit{Metascan-based} methods rely on data that is obtained without interaction from the target system. This can be translated to a scenario in which malware uses Shodan's API to ask whether an IP address is a honeypot before contacting it (e.g. for propagation reasons). We argue that Shodan, Censys and other scanners must introduce limitations to their honeypot identification services. 
From the honeypot deployment and implementation perspective, \acrfull{mtd} techniques could be employed by honeypot implementations to avoid a static IP identification. We also discourage the usage of cloud hosting providers for honeypots based on \acrshort{ics} protocols.

For \textit{probe-based} methods, we suggest that the honeypots are made self-aware and dynamic each time an attack has been detected. Fingerprinting methods can be less effective if the honeypots contain non-static parameters while also choosing selective services periodically. In addition, honeypots rely heavily on protocol emulation libraries. It is important to refer to libraries that are regularly maintained. Furthermore, we suggest making additional tweaks to the references to modify default static responses by comparing the responses to an actual system. Default configurations must be avoided and dynamic configuration based on the attack and the environment is recommended.

\subsection{Shodan Honeyscore}

The Shodan Honeyscore is a proprietary algorithm used to determine whether a crawled instance is a honeypot or not \cite{SHODAN}. Shodan offers an \acrshort{api} that provides a score for \acrshort{ip}s detected as probable honeypots. The score ranges from $[0,0.3,0.5,0.8,1]$, with $0$ denoting that the \acrshort{ip} is not a honeypot and $1$ that it is. The \acrshort{api} also returns the value \textit{NA} when no information is available for a specific IP address.
Since the Honeyscore is not open source, not many conclusions can be derived by examining its output. In fact, it is not disclosed which honeypots can be identified by Shodan's Honeyscore. Nevertheless, we expect that there is some overlap with regard to the fingerprinting techniques used by our framework and Shodan's Honeyscore. 

 \begin{figure}[h!]
    \centering
    \includegraphics[width=0.85\columnwidth]{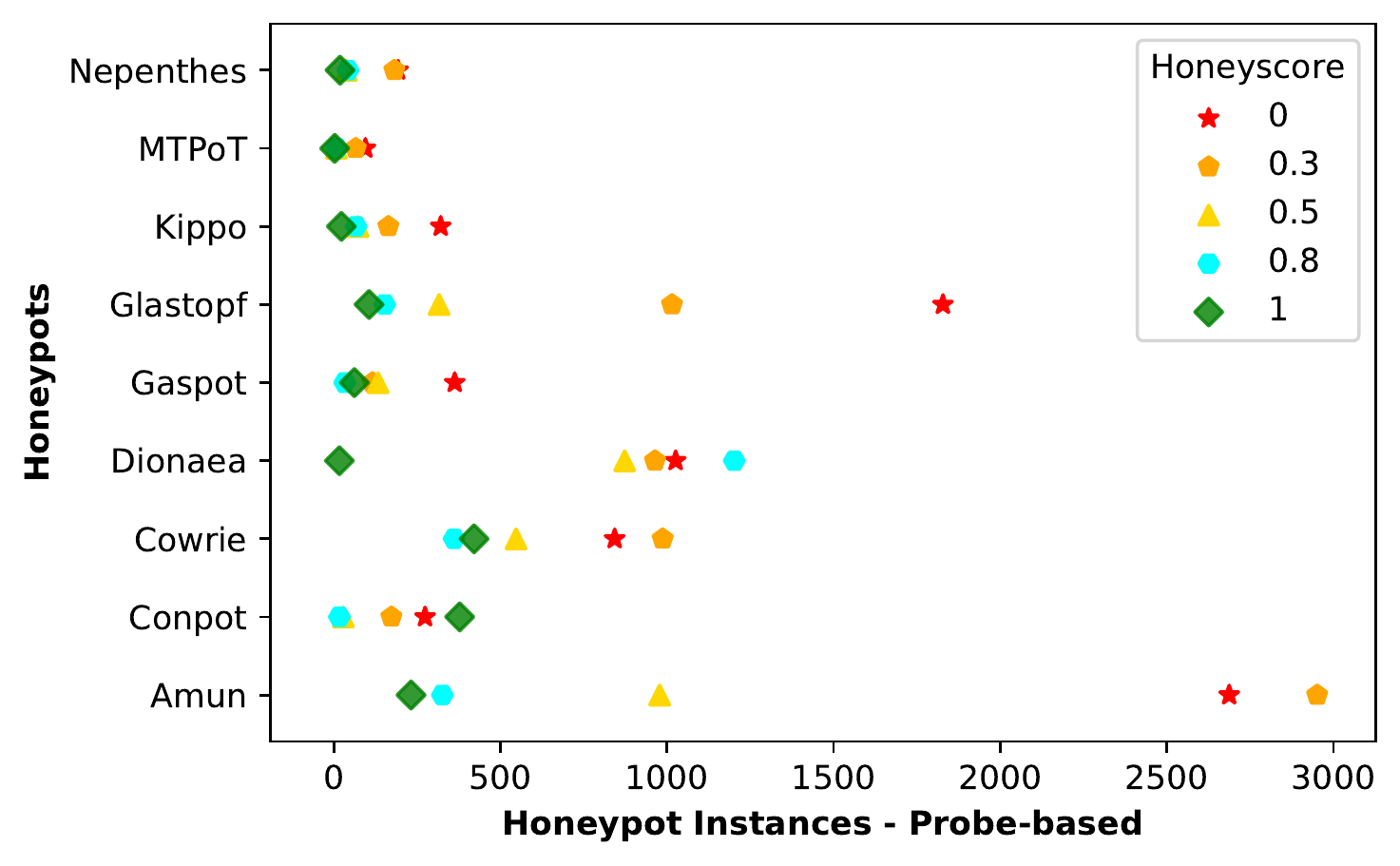}
    \caption{Comparison with Shodan's Honeyscore}
    \label{fig:comparison-hs}
\end{figure}

We fetch the Honeyscore for \textit{all} the honeypot \acrshort{ip}'s determined by our framework and compare the results with Shodan. Figure \ref{fig:comparison-hs} depicts the Honeyscore assigned to honeypot instances detected through our framework (for the combination of both metascan and probe-based results). We observe that Shodan returns $0$ as Honeyscore for many of the \acrshort{ip}s. This suggests that the Honeyscore is not taking into account as many checks as our framework. Moreover, the high deviations observed with regard to Glastopf and Amun suggest that Shodan is not very effective in identifying such honeypots.

\subsection{Limitations}

As discussed in Section \ref{subsec:groundtruth}, the research field of low and medium interaction honeypot fingerprinting has the fundamental limitation that there is no global ground truth knowledge with regard to honeypot deployment. This translates to potential false positives.  
Our work is also influenced by this: while our findings come as the result of multiple stages and checks there may still be cases in which an instance is incorrectly labeled as a honeypot.

The proposed multistage framework leverages multiple checks to determine if the end instance is a honeypot. As part of the failed checks from the framework, 260,000 non-honeypot instances have been detected. While we argue that the majority of these are most likely  vulnerable/misconfigured devices, it might be that some are high-interaction honeypots. Ideally, one could perform manual tests on a sample of these systems by logging into them and attempting to understand the presence of a honeypot environment. However, this would be illegal and therefore we could not perform such an action. Moreover, fingerprinting high-interaction honeypots requires extensive probing and analysis. Hence, this is considered to be out of the scope of this article.

Lastly, while direct comparisons to the state-of-the-art is considered the default evaluation methodology in many fields of cybersecurity, this is not possible in our setting. The combination of the aforementioned ground truth knowledge problem, along with the different time frames of the experiments make direct comparisons unreliable.  
We argue, that our work and results is not competing to the state-of-the-art. This is amplified by the fact that we are dealing with IP addresses and therefore topics such as static vs dynamic IP addresses, \acrfull{nat}, and churn need to be taken into account.

\section{Related Work}
\label{sec:rw}

This section focuses on honeypot-specific fingerprinting research. We note here that besides honeypots, there has been research in the identification of intrusion detection systems and network telescope sensors (e.g. \cite{bethencourt2005mapping,vasilomanolakisprobe2016}). However, we consider this out of the scope of this article. Similarly, we will not discuss here fingerprinting of honeypot-like systems (such as honeytoken identification) \cite{honeytokens2020}. We also note all papers in the SotA exclude high interaction honeypots from their analysis.

\begin{table}[h!]
\centering
\begin{tabular}{|l|l|l|}
\hline
\textbf{Authors \& Year} & \textbf{\begin{tabular}[c]{@{}l@{}}Fingerprinting\\ technique\end{tabular}} & \multicolumn{1}{l|}{\textbf{\begin{tabular}[c]{@{}l@{}} IPv4 scan\end{tabular}}} \\ \hline
Holz et al., 2005 \cite{Holz}  & \begin{tabular}[c]{@{}l@{}}Static command\\ response check\end{tabular} & No \\ \hline
Wang et al., 2010 \cite{wang2010honeypot} & \begin{tabular}[c]{@{}l@{}}Static command\\ response check\end{tabular} & No \\ \hline
Hayatle et al., 2012 \cite{hayatle2012dempster} & \begin{tabular}[c]{@{}l@{}}Static command\\ response check\end{tabular} & No \\ \hline
Aguirre et al., 2014 \cite{aguirre2014new}  & \begin{tabular}[c]{@{}l@{}}Library dependency\\ check, static command\\ response check\end{tabular} & No \\ \hline
Vetterl et al., 2018 \cite{Vetterl2018} & \begin{tabular}[c]{@{}l@{}}Banner check, protocol\\ handshake check,\\ Library dependency\\ check, static command\\ response check\end{tabular} & Yes \\ \hline
Vetterl et al., 2019 \cite{counting}  & \begin{tabular}[c]{@{}l@{}}Banner check,\\ Library dependency\\ check\end{tabular} & Yes \\ \hline
Huang et al., 2019 \cite{huang}  & \begin{tabular}[c]{@{}l@{}}Banner check, static\\ command response\\ check\end{tabular} & No \\ \hline
Morishita et al., 2019 \cite{morishita}  & \begin{tabular}[c]{@{}l@{}}Banner check, http \\ static response\end{tabular} & Yes \\ \hline
Zamiri et al., 2019 \cite{zamiri2019gas} & \begin{tabular}[c]{@{}l@{}}default config,\\static response,\\protocol handshake\end{tabular} & Yes \\ \hline
Papazis et al., 2019 \cite{papazis2019detecting}  & \begin{tabular}[c]{@{}l@{}}Banner check, http\\ static response, static\\ command check\end{tabular} & No \\ \hline
\end{tabular}
\caption{Overview of the related work}
\label{tab:fingerprinting-models}
\end{table}

Techniques for fingerprinting honeypots were first proposed early in 2005 by Holz et al. \cite{Holz}. The authors state that limited simulation and virtualization cause restricted interaction on the honeypot system that leads to fingerprinting possibilities. Holz et al. propose fingerprinting techniques to detect User-mode Linux (UML) kernels by observing the process id's, virtualized environments by analyzing the ping response time, and debuggers by using \textit{ptrace()}. The presented techniques are focused more on fingerprinting at the process and operating system level. This is mainly due to the limited availability of honeypots at the time of research. 

Wang et al. present an approach to detect honeypots in advanced botnet attacks~\cite{wang2010honeypot}. Their work is based on the assumption that security professionals deploying honeypots have a liability constraint; they cannot allow their honeypots to participate in attacks. Hence, botmasters can detect honeypots by checking whether compromised machines in their botnet can successfully send out unmodified malicious traffic. 
This approach is based on monitoring the traffic which is transmitted by the infected system through the bots. For example, the use of \textit{iptables} command on Linux environments to list the port forwarding helps in the identification of honeypots because of outbound traffic rules. This information is transmitted by the bot to the botmaster. 
The authors also present fingerprinting techniques involving ping response time.

Vetterl et al. propose the detection of nine well known open source honeypots by constructing probes to fetch specific data and observe the deviation between the response from actual honeypots~\cite{Vetterl2018}. The deviation is measured as a cosine coefficient. This approach provides a good insight into the state of open source honeypots and their vulnerability to fingerprinting attacks. The methodology is evaluated and the authors identify $7,605$ honeypots on the Internet. In comparison, although our framework employs an approach to observe deviation in responses, we further extend the framework to include additional checks to reduce false positives.

Moreover, Morishita et al. \cite{morishita} propose honeypot fingerprinting through signature-based detection. The authors develop signatures for $15$ open source honeypots offering multiple services. The signatures are then matched against responses obtained through probes and mass-scan engines to determine if the system is a honeypot. The approach is evaluated and the authors detect $19,208$ honeypots. Our approach checks for known honeypot banners returned by the instances, although it does not rely solely on the banner check to flag the instance as a honeypot.

In addition, Zamiri et al. detect GasPot \cite{wilhoitgaspot}, an ATG-based \acrshort{ics} honeypot  through probes designed to fetch information about the default configuration and limited emulation of the protocols  \cite{zamiri2019gas}. The authors study \acrshort{ics} honeypots (specifically of Conpot and GasPot) list features, e.g. limited emulation static responses, and identify the underlying \acrshort{os} to eventually fingerprint them. They perform an Internet-wide scan to detect $17$ GasPot and $240$ Conpot instances. 

Huang et al. probe remote systems and label the response data to train a machine-learning model to classify systems as honeypots \cite{huang}.
The method follows a recursive probing process to obtain featured data for classification. The features include application-layer, network-layer, and system-layer properties. The authors train the model for classification by providing data from known honeypot systems. However, the authors do not classify the responses from widely recognized honeypots like Kippo, Cowrie, or Dionaea.

Lastly, Papazis et al. attempt to exploit some of the virtual network layers implemented in honeypots, using tools like NMap, to fingerprint them \cite{papazis2019detecting}. In addition, they demonstrate the identification of network and service anomalies like link latency and limited emulation that may also lead to honeypot detection. The authors discuss detection vectors for honeypots like Sebek, Artillery, BearTrap, KFSensor, HoneyD, Kippo and Dionaea.

Table \ref{tab:fingerprinting-models} summarizes the fingerprinting related work.
We note that majority of the related work does not evaluate their proposed techniques by performing an active search for honeypots on the Internet. 
This is mainly due to the fact that Internet-wide scanning was not trivial until the emergence of ZMap \cite{zmap}.
That said, the fingerprinting techniques suggested by~\cite{Vetterl2018,morishita,counting,zamiri2019gas} include a thorough evaluation. However, their core limitation is that they focus on a limited number of techniques for fingerprinting.
In this article, we propose a multistage framework that combines probe-based techniques (targeting multiple system layers) with data available from Internet mass-scan search providers to systematically detect honeypots.

\section{Conclusion}
\label{sec:conc}
Honeypots are unique mechanisms for understanding attack methodologies, discovering new attack trends, as well as for early warning systems. In this article, we proposed a framework for honeypot fingerprinting that includes new and SotA components and is able to identify thousands of honeypot instances for \textit{nine} of the most popular honeypot implementations. Our work reduces false positives by the utilization of multiple checks before determining that an instance is a honeypot. Our results also suggest that probe-based fingerprinting techniques are significantly more effective in detecting honeypots than the metascan techniques that utilize third-party systems like Shodan.
We once more highlight that our work is in the direction of improving honeypots rather than arguing against them. With the availability of open honeypot identification APIs , such as Shodan's Honeyscore, it is only a matter of time that we see honeypot-evading malware.
In this context, we contacted both the developers and the administrators of the honeypots to make them aware of potential fingerprinting issues. However, based on the experience of previous work we are not over-optimistic with regard to the patching/updating of such systems. Instead, we argue that novel components must be added in new/old honeypots that are in the direction of \acrlong{mtd}s schemes. We plan to further investigate fingerprinting countermeasures in our future work.

\section*{ACKNOWLEDGMENTS}
This research was supported as part of COM$^3$, an Interreg project supported by the North Sea Programme of the European Regional Development Fund of the European Union.

\bibliographystyle{plain}
\bibliography{main.bbl}  

\appendix

\newpage
\appendix
\section{Multistage Framework for Honeypot Fingerprinting}

Table \ref{tab:hp-banners} shows the banners advertised by honeypots in our evaluation. Most honeypot implementations offer limited banners or custom banners. 
\begin{table}[h]
\centering
\begin{tabular}{|l|l|l|}
\hline
\multicolumn{1}{|c|}{\textbf{Honeypot}} & \multicolumn{1}{c|}{\textbf{Protocol}} & \multicolumn{1}{c|}{\textbf{Banner}} \\ \hline
Kippo & SSH & \begin{tabular}[c]{@{}l@{}}Default: SSH-2.0-OpenSSH\_5.1p1 Debian-5\\ \# SSH-1.99-OpenSSH\_4.3\\ \# SSH-1.99-OpenSSH\_4.7\\ \# SSH-1.99-Sun\_SSH\_1.1\\ \# SSH-2.0-OpenSSH\_4.2p1 Debian-7ubuntu3.1\\ \# SSH-2.0-OpenSSH\_4.3\\ \# SSH-2.0-OpenSSH\_4.6\\ \# SSH-2.0-OpenSSH\_5.1p1 Debian-5\\ \# SSH-2.0-OpenSSH\_5.1p1 FreeBSD-20080901\\ \# SSH-2.0-OpenSSH\_5.3p1 Debian-3ubuntu5\\ \# SSH-2.0-OpenSSH\_5.3p1 Debian-3ubuntu6\\ \# SSH-2.0-OpenSSH\_5.3p1 Debian-3ubuntu7\\ \# SSH-2.0-OpenSSH\_5.5p1 Debian-6\\ \# SSH-2.0-OpenSSH\_5.5p1 Debian-6+squeeze1\\ \# SSH-2.0-OpenSSH\_5.5p1 Debian-6+squeeze2\\ \# SSH-2.0-OpenSSH\_5.8p2\_hpn13v11 FreeBSD-20110503\\ \# SSH-2.0-OpenSSH\_5.9p1 Debian-5ubuntu1\\ \# SSH-2.0-OpenSSH\_5.9\end{tabular} \\ \hline
Cowrie & SSH & Debian GNU/Linux 7 \\ \hline
Cowrie & Telnet & \textbackslash{}xff\textbackslash{}xfd\textbackslash{}x1flogin: \\ \hline
Glastopf & HTTP & Apache httpd \\ \hline
Dionaea & FTP & 220 Welcome to the ftp service \\ \hline
Amun(SMTP) & SMTP & 220 mail\textbackslash{}.example\textbackslash{}.com SMTP Mailserver \\ \hline
Amun(IMAP) & IMAP & a001 OK LOGIN completed \\ \hline
Amun(FTP) & FTP & 220 Welcome to my FTP Server \\ \hline
Conpot & SSH & SSH-2.0-OpenSSH\_6.7p1 Ubuntu-5ubuntu1.3 \\ \hline
Conpot & Telnet & Connected to {[}00:13:EA:00:00:0{]} \\ \hline
Gaspot & ATG & Linux 3.X|4.X \\ \hline
Nepenthes & FTP & 220 ---freeFTPd 1\textbackslash{}.0---warFTPd 1\textbackslash{}.65--- \\ \hline
MTPot & Telnet &\begin{tabular}[c]{@{}l@{}} \textbackslash{}xff\textbackslash{}xfb\textbackslash{}x01\textbackslash{}xff\textbackslash{}xfb\textbackslash{}x03\textbackslash{}xff\textbackslash{}xfc’\textbackslash{}xff\textbackslash{}xfe\textbackslash{}x01\\\textbackslash{}xff\textbackslash{}xfd\textbackslash{}x03\textbackslash{}xff\textbackslash{}xfe\textbackslash{}"\textbackslash{}xff\textbackslash{}xfd’\textbackslash{}xff \textbackslash{}xfd\textbackslash{}x18\textbackslash{}xff\textbackslash{}xfe\textbackslash{}x1f \end{tabular}\\ \hline
\end{tabular}
\caption{Banners advertised by Honeypots (adapted from \cite{Vetterl2018} and \cite{morishita} see \textit{Banner check} in Section \ref{subsub: PBFP})}
\label{tab:hp-banners}
\end{table}


\clearpage
\newpage
Table \ref{tab:hp-http} shows the static content received as HTTP response from honeypots for specific requests. The static responses are either due to limited emulation or due to honeypots being deployed with default configuration.

\begin{table}[h]
\centering
\begin{tabular}{|l|l|l|}
\hline
\textbf{Honeypot} & \multicolumn{1}{l|}{\textbf{\begin{tabular}[c]{@{}l@{}}HTTP \\ Request\end{tabular}}} & \multicolumn{1}{c|}{\textbf{HTTP Response Contents}} \\ \hline
Glastopf & GET / HTTP/1.0 & \begin{tabular}[c]{@{}l@{}}1. \textless{}h2\textgreater{}My Resource\textless{}/h2\textgreater\\ 2. \textless{}h2\textgreater{}Blog Comments\textless{}/h2\textgreater{}\textbackslash{}n \textless{}label for=\textbackslash{}"comment\textbackslash{}"\textgreater{}Please post your comments \\for the blog\textless{}/label\textgreater{}\textbackslash{}n \textless{}br /\textgreater \textbackslash{}n \textless{}textarea name=\textbackslash{}"comment\textbackslash{}" id=\textbackslash{}"comment\textbackslash{}" rows=\textbackslash{}"4\\\textbackslash{}" columns=\textbackslash{}"300\textbackslash{}"\textgreater \textless{}/textarea\textgreater{}\textbackslash{}n \textless{}br /\textgreater{}\textbackslash{}n \textless{}input type=\textbackslash{}"submit\\\textbackslash{}" name=\textbackslash{}"submit\textbackslash{}" id=\textbackslash{}"submit\_comment\textbackslash{}" value=\textbackslash{}"Submit\textbackslash{}" /\textgreater{}\textbackslash{}n\end{tabular} \\ \hline
Amun & GET / HTTP/1.0 & \begin{tabular}[c]{@{}l@{}}\textless{}!DOCTYPE HTML PUBLIC \textbackslash{}"-//IETF//DTD HTML 2\textbackslash{}.0//EN\textbackslash{}"\textgreater{}\textless{}html\textgreater{}\textless{}head\textgreater{}\textless{}title\textgreater{}\\It works!\textless{}/title\textgreater \textless{}/head\textgreater{}\textless{}html\textgreater{}\textless{}body\textgreater{}\textless{}h1\textgreater{}It works!\textless{}/h1\textgreater{}\textless{}br\textgreater{}tim\textbackslash{}.bohn@gmx\textbackslash{}.net\\ \textless{}br\textgreater{}johan83@freenet\textbackslash{}.de\textless{}/body\textgreater{}\textless{}/html\textgreater{}\textbackslash{}n\textbackslash{}n\end{tabular} \\ \hline
Dionaea & GET / HTTP/1.0 & \begin{tabular}[c]{@{}l@{}}\textless{}!DOCTYPE html PUBLIC \textbackslash{}"-//W3C//DTD HTML 3\textbackslash{}.2 Final//EN\textbackslash{}"\textgreater{}\textless{}html\textgreater{}\textbackslash{}n\textless{}title\textgreater\\ Directory listing for  /\textless{}/title\textgreater{}\textbackslash{}n\textless{}body\textgreater{}\textbackslash{}n\textless{}h2\textgreater{}Directory listing for /\textless{}/h2\textgreater{}\textbackslash{}n\end{tabular} \\ \hline
Conpot & \begin{tabular}[c]{@{}l@{}}GET /HTTP/1.0/\\ index.html\end{tabular} & \begin{tabular}[c]{@{}l@{}}1. Last-Modified: Tue, 19 May 1993 09:00:00 GMT\\ 2. Technodrome\\ 3. Mouser Factory\end{tabular} \\ \hline
\end{tabular}
\caption{HTTP Response from Honeypots (see \textit{HTTP Static Response} in Section \ref{subsub: PBFP})}
\label{tab:hp-http}
\end{table}

Table \ref{tab:faile-hp} shows the non-honeypot instances determined at stages in our framework based on honeypot types. Limited emulation in honeypots cause identification at different levels that are determined by the stages in our framework. 
\begin{table}[h]
\centering
\begin{tabular}{|l|c|c|c|c|c|c|c|}
\hline
\textbf{Honeypot} & \multicolumn{1}{l|}{\textbf{Portscan}} & \multicolumn{1}{l|}{\textbf{\begin{tabular}[c]{@{}l@{}}Failed \\ Banner\end{tabular}}} & \multicolumn{1}{l|}{\textbf{\begin{tabular}[c]{@{}l@{}}Failed\\ Static http\\ response\end{tabular}}} & \multicolumn{1}{l|}{\textbf{\begin{tabular}[c]{@{}l@{}}Failed\\ SSL/TLS\\ Certificate check\end{tabular}}} & \multicolumn{1}{l|}{\textbf{\begin{tabular}[c]{@{}l@{}}Failed Protocol\\ Handshake\end{tabular}}} & \multicolumn{1}{l|}{\textbf{\begin{tabular}[c]{@{}l@{}}Failed \\ Library\\ Dependency\\ Check\end{tabular}}} & \multicolumn{1}{l|}{\textbf{\begin{tabular}[c]{@{}l@{}}Not a \\ Honeypot\end{tabular}}} \\ \hline
Kippo & 4361857 & 4324502 & NA & NA & 34887 & 1656 & 4361045 \\ \hline
Cowrie & 4361857 & 4318645 & NA & NA & 37836 & 2100 & 4358581 \\ \hline
Glastopf & 57062712 & 56385819 & 673462 & NA & 0 & 0 & 57059281 \\ \hline
Dionaea & 43944853 & 43890588 & 49963 & 201 & 0 & 0 & 43940752 \\ \hline
Nepenthes & 10391953 & 10391645 & NA & NA & 3 & 0 & 10391648 \\ \hline
Conpot & 29950 & 28693 & NA & NA & 732 & 333 & 29758 \\ \hline
Gaspot & 222593 & 222393 & NA & NA & 0 & 0 & 222393 \\ \hline
MTPot & 2923651 & 2923412 & NA & NA & 0 & 0 & 2923412 \\ \hline
Amun(SMTP) & 6020828 & 6018931 & NA & NA & 0 & 0 & 6018931 \\ \hline
Amun(IMAP) & 4152084 & 4150278 & NA & NA & 0 & 0 & 4150278 \\ \hline
Amun(FTP) & 10391953 & 10389555 & NA & NA & 0 & 0 & 10389555 \\ \hline
Amun(HTTP) & 43944853 & 43942485 & NA & NA & 0 & 0 & 43943476 \\ \hline
\textbf{Total} & \textbf{187809144} & \textbf{186986946} & \textbf{724416} & \textbf{201} & \textbf{73458} & \textbf{4089} & \textbf{187789110} \\ \hline
\end{tabular}
\caption{Non-honeypot encounters by stage. See also Section \ref{subsub:nonhp}.}
\label{tab:faile-hp}
\end{table}

\clearpage
\newpage

Table \ref{tab:hp-ks} denotes the keywords used in Shodan and Censys to retrieve honeypots. The keywords are derived from banners and static content advertised by honeypots. 

\begin{table}[h!]
\centering
\begin{tabular}{|l|l|l|}
\hline
\multicolumn{1}{|c|}{\textbf{Honeypot}} & \multicolumn{1}{c|}{\textbf{Shodan}} & \multicolumn{1}{c|}{\textbf{Censys}} \\ \hline
Glastopf & \textless{}h2\textgreater{}My Resource\textless{}/h2\textgreater{} & 80.http.get.body: "\textless{}h2\textgreater{}My Resource\textless{}/h2\textgreater{}/" \\ \hline
Dionaea & ssl:"Nepenthes" & \begin{tabular}[c]{@{}l@{}}443.https.tls.certificate.parsed.subject.common\_name: \\ "Nepenthes Development Team"\end{tabular} \\ \hline
Conpot & port:"102" product:"Conpot" & 80.http.get.body: "Technodrome" \\ \hline
Nepenthes & \begin{tabular}[c]{@{}l@{}}product:"Nepenthes HoneyTrap \\ fake vulnerable ftpd"\end{tabular} & 21.ftp.banner.banner: "220 ---freeFTPd 1\textbackslash{}.0---warFTPd " \\ \hline
Amun & "220 Welcome to my FTP Server" & \begin{tabular}[c]{@{}l@{}}"80.http.get.body: tim.bohn@gmx.net"\\ 21.ftp.banner.banner: "220 Welcome to my FTP Server"\\ 25.smtp.starttls.banner: "220 mail\textbackslash{}.example\textbackslash{}.com SMTP Mailserver"\\ 143.imap.starttls.banner: "OK LOGIN completed"\end{tabular} \\ \hline
Gaspot & I20100 port: "10001" & "I20100" \\ \hline
\end{tabular}
\caption{Honeypot keywords search (see also paragraph \textit{Keyword Search} in Section \ref{subsub: keyword})}
\label{tab:hp-ks}
\end{table}

Table \ref{tab:hp-sr} shows the static response returned by honeypots for specific commands requested by our probes. Limited emulation or default configuration lead to static response from the honeypots. 
\begin{table}[h!t!]
\centering
\begin{tabular}{|l|l|l|}
\hline
\multicolumn{1}{|c|}{\textbf{Honeypot}} & \multicolumn{1}{c|}{\textbf{Command}} & \multicolumn{1}{c|}{\textbf{Response}} \\ \hline
Conpot & \begin{tabular}[c]{@{}l@{}}S7\_ID\\ station name\\ unit name\end{tabular} & \begin{tabular}[c]{@{}l@{}}88111222\\ "STATOIL STATION"\\ "Technodrome"\end{tabular} \\ \hline
Kippo & \begin{tabular}[c]{@{}l@{}}nano\\ vi\end{tabular} & E558: Terminal entry not found in terminfo \\ \hline
Cowrie & arp & \begin{tabular}[c]{@{}l@{}}IP address       HW type   Flags       HW address            Mask     Device\\ 192.168.1.27  0x1            0x2         52:5e:0a:40:43:c8     *          eth0\\ 192.168.1.1    0x1            0x2         00:00:5f:00:0b:12     *          eth0\end{tabular} \\ \hline
Amun(FTP) & quit & 221 Quit. 221 Goodbye! \\ \hline
Gaspot & I30100 & 9999FF1B \\ \hline

\end{tabular}
\caption{Overview of honeypot static responses. In reference to Section \ref{subsub: scr}}
\label{tab:hp-sr}
\end{table}

\clearpage

Table \ref{tab:hp-scanandtype} provides an overview of the number of honeypot types and instances detected over the three scanning periods.

\begin{table}[h!]
\centering
\begin{tabular}{|c|c|c|c|c|c|}
\hline
\textbf{Honeypot} & \textbf{Scan 1} & \textbf{Scan 2} & \textbf{Scan 3} & \textbf{Total} & \textbf{\begin{tabular}[c]{@{}c@{}}Total\\ (active)\end{tabular}} \\ \hline
Dionaea    & 4101 & 4101 & 4101 & \textbf{4101} & 4101 \\ \hline
Glastopf   & 3431 & 3433 & 3420 & \textbf{3433} & 3420 \\ \hline
Cowrie     & 3276 & 3272 & 3174 & \textbf{3276} & 3174 \\ \hline
Amun(FTP)  & 2398 & 2388 & 2379 & \textbf{2398} & 2379 \\ \hline
Amun(SMTP) & 1897 & 1897 & 1883 & \textbf{1897} & 1883 \\ \hline
Amun(IMAP) & 1806 & 1806 & 1795 & \textbf{1806} & 1795 \\ \hline
Amun(HTTP) & 1377 & 1375 & 1455 & \textbf{1455} & 1455 \\ \hline
Kippo      & 812  & 809  & 796  & \textbf{812}  & 796  \\ \hline
Conpot     & 399  & 751  & 884  & \textbf{884}  & 884  \\ \hline
Nepenthes  & 305  & 302  & 589  & \textbf{589}  & 589  \\ \hline
MTPoT      & 239  & 238  & 215  & \textbf{239}  & 215  \\ \hline
Gaspot     & 200  & 755  & 965  & \textbf{965}  & 965  \\ \hline
\end{tabular}
\caption{Honeypots detected per scan}
\label{tab:hp-scanandtype}
\end{table}

\clearpage





\section{Framework specific checks and pipeline}

Algorithm \ref{algo:cert-check}, represents the pseudo-code block that checks an instance for Dionaea's default certificate parameters. In lines $3$-$5$, the algorithm retrieves the certificate from the web server by accepting the \acrshort{ip} and port of the instance and checks for common attributes subject organization, country, and issuer. These attributes have static values assigned by the honeypot developers.  In steps $7$-$9$, the algorithm checks if the values match the Dionaea honeypot certificate's static values. Upon match, the algorithm returns that the instance is a Dionaea honeypot.

\begin{algorithm}[]
  \caption{Certificate Check}
\label{algo:cert-check}
  \footnotesize
  \DontPrintSemicolon
  \SetKwInOut{Input}{input}\SetKwInOut{Output}{output}
  \Input{\var{ip}, \var{port}} 
  \Output{isDionaea \tcc*[f] {True if certificate from Dionaea}}
  \Begin{
    \nl \proc{checkCert}{\var{ip},\var{port}}\;
    \nl $\var{isDionaea} = \var{false}$\;  
    \nl $\var{cert} =  \proc{ssl.get\_server\_cert}{\var{ip},{\var{port}}}$\;
    \nl $\var{X509} = \var{Crypto.X509.load\_cert(cert)}$\; 
    \nl $\var{org} = \var{X509.subject.org}$\; 
         \nl\If{\var{cert}}{       
        \nl \If{\var{org}= {\var{"dionaea.carnivore.it"}}}{
        \nl $\var{isDionaea} = \var{true}$\;
        }
        \nl $\var{Return isDionaea}$\;
      }\nl $\var{\textbf{end}}$
      }
\end{algorithm}

Algorithm \ref{algo: metapipe} shows the checks done in the Metascan-based pipeline. The algorithm checks or open ports, and performs keyword based check to list instances that match the static content delivered by honeypots. Furthermore, additional checks like \acrshort{fqdn} and cloud hosting checks are performed for determining specific honeypot types.

Algorithm \ref{algo: metsearch} represents the metascan search performed to determine instances with specific ports exposed to the Internet. The algorithm performs a search on Shodan and Censys mass scan engines for specific ports that are open on honeypots in our test.
\begin{algorithm}[]
  \caption{Metascan-based Pipeline}
    \label{algo: metapipe}
  \footnotesize
  \DontPrintSemicolon

  \SetKwInOut{Input}{input}\SetKwInOut{Output}{output}
  \Input{\var{ports} \tcc*[f] {ports}}
  \Output{findHoneypot \tcc*[f] {Honeypots on the Internet}}

  \Begin{
    \nl $\var{ip} \gets \proc{metasearch}{\var{ports}}$\;   
    \tcc*[f] {Shodan and Censys Search} \\
    \nl\ForEach{\var{ip}}{
             
        \nl $\var{kw}=\proc{keywordSearch}{\var{ip}}$\;
        \nl\If{kw}{
                \nl \ForEach{\var{ip}}{}{
                   \nl \If{\proc{checkfqdn}{\var{ip}}}{
                            \nl return $\var{hasFqdn}=false$\;
                             }
                      \nl  \If{\var{!hasfqdn} \& \var{port=502|102}}{
                    \nl       \If{\proc{cloudCheck}{\var{ip}}}{
                    \nl     return $\var{isHoneypot}=true$\;  
                           
                           }
                           \nl \var{isHoneypot}
                                 
                        }     
                \nl    \If{\var{!hasfqdn}}{
                     \nl \If{\proc{isResearch}{\var{ip}}}{
                     \nl return $\var{isResearch}=true$\;       
                           }
                       \nl \var{isHoneypot}
                      }
                    }
                \nl endIf
        
        } \nl endFor
      
    }\nl end
  }
  
\end{algorithm}

\begin{algorithm}[]
  \caption{Metascan Search}
 \label{algo: metsearch}
  \footnotesize
  \DontPrintSemicolon

  \SetKwInOut{Input}{input}\SetKwInOut{Output}{output}
  \Input{\var{port} \tcc*[f] {search parameter}}
  \Output{instanceIP \tcc*[f] {Instances with open ports}}

  \Begin{
    \nl $\var{instances[]}=null$\;
    \nl \proc{shodanSearch}{\var{port}}\; 
    \nl \ForEach{ip}{
        \nl  $\var{instances[].append}(\var{ip}, \var{port} )$\;
        \nl return $\var{instances[]}$\;
        \nl endFor
        } 
     \nl \proc{censysSearch}{\var{port}} \;
         \nl \ForEach{ip}{
        \nl  $\var{instances[].append}(\var{ip}, \var{port})$\;
        \nl return $\var{instances[]}$\;
        \nl endFor
        } 
\nl end
}
\end{algorithm}

Lastly, algorithm \ref{algo:ph} represents the Protocol Handshake check procedure described in Section \ref{subsub:ph}. The algorithm checks for a deviated response from the instances for specific negotiation parameters and response based on the port and the service of the instance. 
\newcommand*{\escape}[1]{\texttt{\textbackslash#1}}
\begin{algorithm*}[h!]
  \caption{Protocol Handshake Check}
  \label{algo:ph}

  \footnotesize
  \DontPrintSemicolon

  \SetKwInOut{Input}{input}\SetKwInOut{Output}{output}
  \Input{\var{instance[]},  \tcc*[f] {Instance [ip, port, protocol, isDeviated]}}
  \Output{\var{instance[isDeviated]} \tcc*[f] {Handshake is deviated}}

  \Begin{
    \nl \proc{CheckHandshake}{\var{instance[]}}\;
    \tcc*[f] {For each instance} \\[-\baselineskip]
    \nl\ForEach{\var{ip in Instance[]}}{
          \nl\If{\var{port=22/2222 \& protocol="SSH"}}{       
             \nl \proc{request}{{"SSH-2.0-OpenSSH\escape{n}\escape{n}\escape{n}\escape{n}\escape{n}\escape{n}\escape{n}\escape{n}\escape{n}\escape{n}"}} \;
              \nl if \var{response} = "bad Packet length" or "protocol mismatch"\;
             \nl return $\var{Instance[isDeviated]}=true$\;
             \nl \textbf{else} \proc{request}{{"SSH-2.0-OpenSSH\_6.0p1 Debian-4+deb7u2\escape{n}"}} \;
             \nl if \var{response} = "protocol mismatch\escape{n}"\;
             \nl return $\var{Instance[isDeviated]}=true$\;
            } 
           \nl \textbf{else}\If{\var{port=102 \& protocol="S7"}}{       
             \nl \proc{request}{H,0300002102f08032070000...} \;
              \nl if \var{response} = "0x32"\;
             \nl return $\var{Instance[isDeviated]}=true$\;
            } 
            \nl \textbf{else}\If{\var{port=502 \& protocol="Modbus"}}{       
             \nl \proc{request}{function\_code:None, slave\_id:0, request:0000000000050..} \;
              \nl if \var{session.state} = "Disconnected"\;
             \nl return $\var{Instance[isDeviated]}=true$\;
            } 
        \nl \textbf{else}\If{\var{port=25 \& protocol="SMTP"}}{       
             \nl \proc{request}{PASS:Test} \;
              \nl if \var{response} = "220 OK"\;
             \nl return $\var{Instance[isDeviated]}=true$\;
            }   
        \nl \textbf{else}\If{\var{port=143 \& protocol="IMAP"}}{       
             \nl \proc{request}{RCPT TO:TEST} \;
              \nl if \var{response} = "221 Bye Bye"\;
             \nl return $\var{Instance[isDeviated]}=true$\;
            }     
         \nl \textbf{else}\If{\var{port=21 \& protocol="FTP"}}{       
             \nl \proc{request}{ftp (ip)} \;
              \nl if \var{packet.windowSize}=4096 \& \var{session.disconnect.timeout}=45 \;
             \nl return $\var{Instance[isDeviated]}=true$\;
            } 
        \nl \textbf{else}\If{\var{port=23/2323 \& protocol="Telnet"}}{       
             \nl \proc{request}{telnet (ip)} \;
              \nl if \var{packet.response}= "You have connected to the telnet server" \;
             \nl return $\var{Instance[isDeviated]}=true$\;
            }
         \nl \textbf{else}\If{\var{port=80/8080/443/8443 \& protocol="HTTP/HTTPS"}}{       
             \nl \proc{request}{GET /HTTP/1.0 (ip)} \;
              \nl if \var{response.packet.header.server}= "nginx" or "Apache/1.3.29" or "BaseHTTP/0.3 Python/2.5.1" or "Microsoft-IIS/5.0" \;
             \nl return $\var{Instance[isDeviated]}=true$\;
            }
            
    }
  }
\end{algorithm*}

\end{document}